\documentclass[twocolumn]{aastex63}

\usepackage{bm}
\usepackage{enumitem}
\setlist{nosep}

\newcommand{\simname}[1]{\texttt{#1}}

\received{October 29, 2020}
\revised{January 8, 2021}
\accepted{January 10, 2021}
\submitjournal{ApJ}

\shorttitle{Low Metallicity PPDs}
\shortauthors{Kadam et al.}
\graphicspath{{./}{figures/}}

\begin{document}

\title{Eruptive Behavior of Magnetically Layered Protoplanetary Disks in Low Metallicity Environments}

\correspondingauthor{Kundan Kadam}
\email{kkadam@uwo.ca}

\author[0000-0001-8718-6407]{Kundan Kadam}
\affiliation{University of Western Ontario, Department of Physics and Astronomy, London, Ontario, N6A 3K7, Canada}
\affiliation{Konkoly Observatory, Research Center for Astronomy and Earth Sciences, Konkoly-Thege Mikl\'{o}s \'{u}t 15-17, 1121 Budapest, Hungary}

\author[0000-0002-6045-0359]{Eduard Vorobyov}
\affiliation{Department of Astrophysics, The University of Vienna, A-1180 Vienna, Austria}
\affiliation{Research Institute of Physics, Southern Federal University Rostov-on-Don, 344090 Russia}

\author[0000-0001-7157-6275]{\'{A}gnes K\'{o}sp\'{a}l}
\affiliation{Konkoly Observatory, Research Center for Astronomy and Earth Sciences, Konkoly-Thege Mikl\'{o}s \'{u}t 15-17, 1121 Budapest, Hungary}
\affiliation{Max Planck Institute for Astronomy, K\"onigstuhl 17, D-69117 Heidelberg, Germany}
\affiliation{ELTE E\"otv\"os Lor\'and University, Institute of Physics, P\'azm\'any P\'eter s\'et\'any 1/A, 1117 Budapest, Hungary}



\begin{abstract}
A protoplanetary disk typically forms a dead zone near its midplane at the distance of a few au from the central protostar.
Accretion through such a magnetically layered disk can be intrinsically unstable and has been associated with episodic outbursts in young stellar objects.
We present the first investigation into the effects of low metallicity environment on the structure of the dead zone as well as the resulting outbursting behavior of the protoplanetary disk.
We conducted global numerical hydrodynamic simulations of protoplanetary disk formation and evolution in the thin-disk limit.
The consequences of metallicity were considered via its effects on the gas and dust opacity of the disk, the thickness of the magnetically active surface layer, and the temperature of the prestellar cloud core.
We show that the metal poor disks accumulate much more mass in the innermost regions, as compared to the solar metallicity counterparts.
The duration of the outbursting phase also varies with metallicity -- the low metallicity disks showed more powerful luminosity eruptions with a shorter burst phase, which was confined mostly to the early, embedded stages of the disk evolution.
The lowest metallicity disks with the higher cloud core temperature showed the most significant differences.
The occurrence of outbursts was relatively rare in the disks around low mass stars and this was especially true at lowest metallicities.
We conclude that the metal content of the disk environment can have profound effects on both the disk structure and evolution in terms of episodic accretion.  

\end{abstract}



\section{Introduction} 
\label{sec:intro}

The general picture of low-mass star formation suggests that, as a result of the conservation of angular momentum, a pre-main sequence star inevitably forms a surrounding flattened accretion disk.
Planetary systems are born within the environment of such protostellar (younger and relatively massive) and protoplanetary disks (PPDs).
It is widely accepted that the disk accretion in such systems, also called young stellar objects (YSOs), is not steady but time-dependent, with low mass accretion rates punctuated by episodes of powerful eruptions \citep[e.g., see][and references therein]{Audard2014}.
This picture is also supported by the surveys of protostars, which suggest that the luminosity of YSOs is consistently about an order of magnitude lower than that expected from steady accretion of mass \citep[][]{Eisner2005,Dunham2010}. 
Sudden and luminous accretion events known as FUor ($\sim100 L_\odot$, $\sim100$ years, more resembling embedded Class I YSOs) and EXor ($\sim10 L_\odot$, $\sim1$ year, spanning both Class I/II) type eruptions have also been directly observed in YSOs \citep[][and references therein]{HK1996,Herbig2008,Audard2014}.
On average about 10\% of the final stellar mass of a low mass star is thought to be cumulatively accreted during episodic outbursts, which can reach as high as 35\% in extreme cases \citep{DV12}.
{Growing evidence suggests that} accretion bursts may occur during the formation of massive stars as well \citep[][]{CoG2017,Meyer17,Magakian2019}.
The magnitude of these eruptions is large enough to have substantial effects on disk dynamics, chemistry, mineralogy, dust properties and snow lines, all of which have significant consequences for planet formation.

After its mass, metallicity of a star (i.e., content of species heavier than helium) has the most significant effect on its structure and evolution \citep[e.g.,][]{HKT}.
Hence it is reasonable to assume that the effects of metal content for a PPD may be similarly profound.
However, the observations related to low metallicity PPDs remain poor and controversial, especially in their early stages.  
The dust-to-gas ratio in a PPD is difficult to measure observationally and it is generally considered to be proportional to the metallicity of the host star \citep{Murray01,ErcolanoClarke10}.
The mass accretion rate in young low metallicity stars is observed to be higher than corresponding solar metallicity systems
\citep{Spezzi12,DeMarchi13}.
The inner disk (up to a few au) fraction of low-metallicity YSOs is significantly smaller, suggesting that these disks are dispersed at an earlier stage and have a shorter lifetime due to enhanced photoevaporation -- about 1 Myr as compared to 5 Myr for solar metallicity counterparts \citep{Yasui10,ErcolanoClarke10}.
The frequency of giant planets is strongly correlated with the host star's metallicity, possibly because of an enhanced rate of planetesimal formation \citep{Gonzalez97,FischerValenti05, Johansen09}. 
Due to the difficulties related to time-domain astronomy at large distances, even less is known about episodic accretion in low metallicity environments.
All of the outbursting YSOs discovered so far have been found in star-forming regions within the immediate solar neighborhood, having approximately solar metallicity. 
Once the eruption begins, the stellar photosphere is not visible and hence we cannot measure the metallicity of an ongoing event.

From a theoretical point of view, the episodic accretion in solar metallicity disks has been studied extensively with the help of several numerical models \citep{BL1994,Armitage2001,BB1992,VB2005,DS2010}.
Most relevant for this paper, \cite{Gammie1996} showed that a typical PPD forms a magnetically ``dead zone'' at its midplane due to insufficient ionization to sustain magnetorotational instability (MRI) turbulence.
As a consequence, the accretion occurs only through the active surface layers, with the dead zone forming an effective bottleneck in the mass and angular momentum transport.
Accretion through such a layered disk structure can be intrinsically unsteady, and sudden activation of MRI in the dead zone can give rise to ``MRI-bursts'' \citep{Armitage2001,Zhu10a, Kadam20}.
In the early stages of its evolution, a PPD can also be prone to vigorous gravitational instability (GI), where the disk self-gravity forms large-scale spirals as well as gravitationally bound clumps or fragments \citep{Toomre64, Kratter16}. 
These clumps usually migrate inward on dynamical timescales and their accretion onto the central protostar can trigger luminosity outbursts \citep{VB2015, Meyer17, Zhao18}.
Both of these mechanisms -- MRI-bursts and clump accretion -- have been shown to be consistent with the observational constraints on the longer duration, FUor-type outbursts.

When considering low metallicity environments, numerical hydrodynamic simulations have been carried out to study primarily its effects on disk gravitational fragmentation and possible formation of gas giant planets through this process.
Evolution of inviscid models suggested that a factor of 10 increase or decrease in metallicity does not affect the disk evolution significantly \citep{Boss02}.
\cite{Cai06} supported these findings - that the outcome of hydrodynamic simulations are somewhat insensitive to the metallicity. However, in contradiction with \cite{Boss02}, the cooling times in these simulations were too long for the disks to fragment and form clumps.
In steady state $\alpha$ disk models, the low metallicity disks tend to be more GI unstable \citep{TanakaOmukai14}. 
With hydrodynamic simulations including prestellar clouds collapse and dust physics, \cite{Vorobyov20} showed that metal poor disks are also GI unstable.
The low metallicity models in this study showed that the duration of the burst phase  of the disk due to clump accretion was much shorter than their solar metallicity counterpart.
In all of the low metallicity studies so far, the underlying models assumed a fixed \cite{SS73} $\alpha$ parameter for representing the turbulent viscosity in the disk.
The dead zone in the disk was thus neglected, unless the evolution of the entire disk was carried out with a reduced viscosity ($\alpha$ parameter less than the canonical value of $10^{-2}$).  
The innermost disk region extending a few au, which is the most relevant in terms of observed FUor outbursts \citep{Zhu07}, was also excluded from the computational domain.

In this paper, we present the effects of low metallicity environment on the structure and evolution of PPDs with a focus on the dead zone in the inner disk as well as on the episodic accretion.
We use numerical hydrodynamic simulations of PPDs in the think-disk limit for these investigations. 
The simulations start with the collapse phase of the molecular cloud core so that the initial conditions and mass loading of the disk are accurately reproduced.
The inner boundary of the computational domain is set at 0.42 au so that the sub-au scale behavior of the disk can be captured.
Note that the extent of our simulations, spanning distances from sub-au scale to the radius of the parent molecular cloud as well as temporal evolution over the first 0.7 Myr, push the limits of two-dimensional hydrodynamic models.
The dust grain content of a PPD increases with metallicity which affects several physical processes.
The dust dominates the opacity of a disk and thus controls its cooling properties and thermal equilibrium \citep{Lodato08}.
We model a lower metallicity as a lower dust-to-gas ratio by scaling down the gas and dust opacities in proportion \citep{Boss02,Cai06}.
The low metal content offers a lower opacity to the stellar X-ray radiation and can thus allow a larger column density of the disk to be ionized \citep{Hartmann06}.
This can increase the thickness of the MRI active surface layers.
During the core collapse phase, the decrease in dust continuum emission in the low metallicity environment can lead to a higher background gas temperature \citep{Vorobyov20}.
We take into account the consequences of reduced metallicity in terms of all three effects -- reduced opacity, increased active layer thickness, and increased cloud core temperature (Section \ref{sec:methods}).
We find that the low metallicity disks form progressively more restrictive dead zones in the inner disk and, as a consequence, are more massive (Section \ref{subsec:opacity}).
The length of the phase during which the disk is prone to outbursts decreases at low metallicities, especially when the effects of inefficient cooling are taken into account (Sections \ref{subsec:opacity}, \ref{subsec:sigmaTc}). 
The outbursts are relatively infrequent in the disks around lower mass stars, and may be entirely suppressed at lowest metallicities (Section \ref{subsec:ML}).
An individual outburst in a low metallicity disk is typically more luminous and short-rising as compared to its solar metallicity counterpart (Section \ref{subsec:burst}).  
We conclude that the metal poor environment has a significant effect on the eruptive behavior of PPDs -- the disks tend to be more massive in the innermost regions and exhibit shorter duration of the burst phase, confining it to the initial embedded phases of the YSO.

\section{Model Description and Initial Conditions}
\label{sec:methods}

In this section we will describe the hydrodynamic model used for studying the metallicity effects described in Section \ref{sec:results}.
We will also elaborate on the initial conditions for the simulations and the physical reasons behind the chosen set of model parameters considered in this paper.

The model of PPD formation and evolution is based on the \simname{FEoSaD} (Formation and Evolution Of a Star And its circumstellar Disk) code, where the full set of numerical hydrodynamic equations are solved in the thin-disk limit with a cylindrical geometry \citep{VB2010, VB2015, Vorobyov2018}.
\simname{FEoSaD} has several physics modules which can be turned on depending on the problem at hand, in order to find a balance between the gained insights and the computational cost.
The following hydrodynamic equations of mass, momentum, and energy transport are solved
\begin{equation}
\frac{\partial\Sigma}{\partial t} =-\bm{\nabla}_{p} \cdot\left(\Sigma \bm{v}_{p}\right),
\label{eq:mass}
\end{equation}
\begin{equation}
\frac{\partial}{\partial t}\left(\Sigma \bm{v}_{p}\right)+\left[\bm{\nabla} \cdot \left(\Sigma 
\bm{v}_{p}\otimes \bm{v}_{p}\right)\right]_{p}=-\bm{\nabla}_{p} P+\Sigma \bm{g}_{p}+\left(\bm{\nabla}\cdot
\bm{\Pi}\right)_{p},
\label{eq:momentum}
\end{equation}
\begin{equation}
\frac{\partial e}{\partial t}+\bm{\nabla}_{\mathrm{p}}\cdot\left(e v_{p}\right)=-P\left(\bm{\nabla}_{p} \cdot v_{p}\right)-\Lambda+\Gamma+\left(\bm{\nabla} \cdot v\right)_{pp'}: \bm{\Pi}_{pp'}.
\label{eq:energy}
\end{equation}
Here the subscripts $p$ and $p'$ refer to the planar components in polar coordinates $(r,\phi)$, $\Sigma$ is the surface mass density, $e$ is the internal energy per unit area, and $P$ is the vertically integrated gas pressure. 
Amongst non-scalars, $\bm{v}_{p}=v_{\mathrm{r}}\hat{\bm{r}}+v_{\mathrm{\phi}}\hat{\bm{\phi}}$
is the velocity in the disk plane, $\bm{g}_{p}=g_{\mathrm{r}} \hat{\bm{r}} + g_{\mathrm{\phi}} \hat{\bm{\phi}}$ is the gravitational acceleration in the disk plane and 
$\bm{\nabla}_{\mathrm{p}}=\hat{\bm{r}}\partial/\partial r+\hat{\bm{\phi}}r^{-1}\partial/\partial\phi$
is the gradient along the planar coordinates of the disk.
The ideal equation of state, $P=(\gamma-1)e$ with $\gamma=7/5$, is used for calculating the gas pressure.
The viscous stress tensor,
\begin{equation}
\mathbf{\Pi} = 2 \Sigma \nu \left( \bm{\nabla} v - \frac{1}{3}(\bm{\nabla} \cdot v) {\bf e}  \right),
\label{eq:stresstensor}
\end{equation}
accounts for turbulent viscosity in the disk, where $\nu$ is the kinematic viscosity, $\nabla v$ is a symmetrized velocity gradient tensor and ${\bf e}$ is a unit tensor.

The cooling and heating rates $\Lambda$ and $\Gamma$, respectively, in Equation \ref{eq:energy} are based on the analytical solution of the radiation transfer equations in the vertical direction \citep{Dong2016}.
The cooling function per surface area of the disk is expressed as
\begin{equation}
\Lambda=\frac{8\tau_{\rm P} \sigma T_{\rm mp}^4 }{1+2\tau_{\rm P} + 
\frac{3}{2}\tau_{\rm R}\tau_{\rm P}},
\end{equation}
where  $\sigma$ the Stefan-Boltzmann constant, $\tau_{\rm P}$  and $\tau_{\rm R}$ are the Planck and  Rosseland optical depths to the disk midplane, and  $\kappa_P$ and $\kappa_R$ are the Planck and Rosseland mean opacities, respectively.
The opacities are calculated for physical conditions typical of PPDs which include both dust and gas components \citep{Semenov2003}.
The heating rate is given by
\begin{equation}
\Gamma=\frac{8\tau_{\rm P} \sigma T_{\rm irr}^4 }{1+2\tau_{\rm P} + \frac{3}{2}\tau_{\rm R}\tau_{\rm P}},
\label{eq:heating}
\end{equation}
where $T_{\rm irr}$ is the irradiation temperature at the disk surface
\begin{equation}
T_{\rm irr}^4=T_{\rm bg}^4+\frac{F_{\rm irr}(r)}{\sigma},
\label{eq:Tirr}
\end{equation}
where $T_{\rm bg}$ is the temperature of the background black-body irradiation, $F_{\rm irr}(r)$ is the radiation flux absorbed by the disk surface at radial distance $r$ from the central star. 
The latter is calculated as 
\begin{equation}
F_{\rm irr}(r)= \frac{L_\ast}{4\pi r^2} \cos{\gamma_{\rm irr}},
\label{eq:Firr}
\end{equation}
where $\gamma_{\rm irr}$ is the incidence angle of radiation arriving at the flaring disk surface at radial distance $r$. 
The stellar luminosity $L_\ast$ is the sum of the accretion luminosity and the photospheric luminosity.
The accretion luminosity, $L_{\rm \ast,accr}=(1-\epsilon) G M_\ast \dot{M}/2 R_\ast$, was generated from the accreted gas, where the fraction of accretion energy absorbed by the star ($\epsilon$) was set to 0.05.  
The photospheric luminosity, $L_{\rm \ast,ph}$, was due to both the gravitational compression and the deuterium burning in the stellar interior \citep{VB2010}. 
The stellar mass, $M_\ast$, and accretion rate onto the star, $\dot{M}$, are determined using the amount of mass passing through the inner computational boundary. 
The properties of the forming protostar ($L_{\rm \ast,ph}$ and radius $R_\ast$) are calculated using the pre-main-sequence stellar evolution tracks of \citet{DAntona97}. The effects of metallicity on stellar evolution were not taken into account in this study.

The magnetorotational instability (MRI) is considered to be the primary driver in producing turbulent viscosity in PPDs \citep{HGB1995,Turner2014}.
The disk material needs to be sufficiently ionized for MRI to operate.
The shearing Keplerian motion of the gas coupled with the field causes turbulence and angular momentum transport.
In a typical PPD the midplane temperature outside of about 1 au radius is not high enough to sustain collisional ionization \citep{Armitage2011}.
Galactic cosmic rays are considered to be the major source of ionization at a few au, which penetrate an approximately constant column density of the  gas \citep{UN1981}.
The accretion disk thus forms a layered structure -- accretion occurs only through the sufficiently ionized, MRI-active surface layers and a magnetically dead zone is formed at the midplane \citep{Gammie1996}.

We model the accretion through a layered PPD with an effective and adaptive \citet{SS1973} $\alpha$-parameter \citep{Bae2014, Kadam19}.
The kinematic viscosity is parametrized as, $\nu=\alpha_{\rm eff} c_{\rm s} H$, where $c_{\rm s}$ is the sound speed and $H$ is the vertical scale height of the disk, which is computed assuming a local hydrostatic equilibrium taking disk self-gravity into account \citep{VB09}. The $\alpha_{\rm eff}$ is given by 
\begin{equation}
\alpha_{\rm eff}=\frac{\Sigma_{\rm a} \alpha_{\rm a} + \Sigma_{\rm d} \alpha_{\rm d}}{\Sigma_{\rm a} + \Sigma_{\rm d}},
\label{eq:alpha}
\end{equation}
where $\Sigma_{\rm a}$ is the gas surface density of the MRI-active surface layer of the disk and $\Sigma_{\rm d}$ is the gas surface density of the magnetically dead layer at the disk midplane. 
Note that the total gas surface density is then $\Sigma = 2 \times (\Sigma_{\rm a} + \Sigma_{\rm d})$.
The parameters $\alpha_{\rm a}$ and $\alpha_{\rm d}$ are proportional to the strength of turbulent viscosity in the MRI-active and MRI-dead layers of the disk, respectively. 
We set a canonical value of $\alpha_{\rm a}=0.01$ for the magnetically active region. 
However, there is some uncertainty in $\alpha_{\rm a}$ and we will explore the possibility of a higher value in the future, as suggested by recent magnetohydrodynamic simulations \citep{Zhu18RMHD}.
The parameter $\alpha_{\rm d}$ is defined as 
\begin{equation}
\alpha_{\rm d} = \alpha_{\rm MRI,d} + \alpha_{\rm rd},
\label{eq:alpha_d}
\end{equation}
where
\begin{equation}
\alpha _{\rm MRI,d} =
\left\{
\begin{array}{c}
\alpha_{\rm a}, \,\, \mathrm{if} \,\, T_{\rm mp} > T_{\rm crit}   \\
0, \,\,\, \mathrm{otherwise},
\label{eq:MRI_d}
\end{array}
\right. 
\end{equation}
where $T_{\rm crit}$ is the MRI activation temperature and $T_{\rm mp}$ is the disk midplane temperature.
The viscosity thus sharply rises to the fully MRI-active value above $T_{\rm crit}$. 
This models the almost exponential increase in disk ionization fraction due to the thermal effects, such as dust sublimation and ionization of alkali metals.  
The dead zone in a PPD can have a small but non-zero residual viscosity due to hydrodynamic turbulence driven by the Maxwell stress in the disk active layer \citep{Okuzumi11,Bae2014}.
We set this residual viscosity using
\begin{equation}
\alpha_{\rm rd} = \min \left( 10^{-5} , \alpha_{\rm a} \frac{\Sigma_{\rm a}}{\Sigma_{\rm d}} \right).
\label{eq:alphaRD}
\end{equation}
As the non-zero $\alpha_{\rm rd}$ is due to turbulence propagating from the active layer down to the disk midplane, the above expression ensures that the accretion in the dead zone cannot exceed that of the active layers. 

The numerical simulations start from the gravitational collapse of a starless molecular cloud core.
The protostar is formed within the inner computational boundary of the disk while the envelope continues to accrete inside the centrifugal radius during the embedded phase.
The initial surface density and angular velocity profiles of the cloud core are derived from an axisymmetric core compression where the angular momentum remains constant and magnetic fields are expelled due to ambipolar diffusion \citep{Basu1997}
\begin{equation}
   \Sigma = \frac{r_0 \Sigma_0}{\sqrt{r^2+r_0^2}},
 \end{equation}
 \begin{equation}
   \Omega = 2 \Omega_0 {\left( \frac{r_0}{r} \right)}^2  \left[  \sqrt{1+ {\left( \frac{r}{r_0} \right)}^2}  -1 \right], 
   \end{equation}
where $\Sigma_0$ and $\Omega_0$ are the maximum values at the center of the core and $r_0$ is the radius of the central plateau proportional to the thermal Jeans length.
The initial cloud cores were constructed such that the ratio of the rotational to the gravitational energy, $\beta$, was approximately $0.14\%$. 
This value is consistent with the observations of pre-stellar cores \citep{Caselli2002} and was chosen to be relatively low, in order to inhibit the formation of self-gravitating clumps during the early, massive phases of the disk evolution \citep{Vorobyov13}. 
This was done to avoid the possible interference with another accretion burst model caused by clump infall, which was shown to operate in solar- and low-metallicity environments \citep{Vorobyov20}.

\begin{table*}
\caption{List of Simulations}
\label{table:sims}
\begin{tabular}{|l|}
\hline
\begin{tabular}{p{2cm}p{1.5cm}p{2.0cm}p{1.5cm}p{2.5cm}p{2.5cm}}
\hspace{-1.2cm} Model Name & \hspace{-0.4cm} ${ M_{ \rm core} (M_{\odot}) }$  &  ${ \Sigma_{\rm a}}$ (g~cm$^{-2}$)  & \hspace{-0.5cm}  $ T_{\rm core} {\rm (K) } $ & \hspace{-0.5cm} ${ Z / Z_{\odot}}$&    \hspace{-0.4cm}  Included effects \\
  \end{tabular}\\ \hline
\begin{tabular}{l}
$\kern-\nulldelimiterspace\left.
  \begin{tabular}{p{3.2cm}p{1.5cm}p{1.5cm}p{1.5cm}p{1.5cm}}
\hspace{-0.7cm}\simname{MS\_fid}   & 1.152   & 100 & 15 & 1.0  \\ 
\end{tabular}\right\}$ Fiducial solar mass model
\\ 
\end{tabular}  
\\ \hline
\begin{tabular}{l}
$\kern-\nulldelimiterspace\left.
  \begin{tabular}{p{3.2cm}p{1.5cm}p{1.5cm}p{1.5cm}p{1.5cm}}
\hspace{-0.7cm}\simname{MS\_Z0.1}   & 1.152   & 100 & 15 & 0.1  \\ 
\hspace{-0.7cm}\simname{MS\_Z0.02}   & 1.152   & 100 & 15 & 0.02  \\
  \end{tabular}\right\}$ $ \kappa $ only
\end{tabular}  
\\
\hline
\hline
\begin{tabular}{l}
$\kern-\nulldelimiterspace\left.
  \begin{tabular}{p{3.2cm}p{1.5cm}p{1.5cm}p{1.5cm}p{1.5cm}}
\hspace{-0.7cm}\simname{MS\_$\Sigma$a\_Z0.1}   & 1.152   & 200 & 15 &0.1  \\ 
\hspace{-0.7cm}\simname{MS\_$\Sigma$a\_Z0.02}   & 1.152   & 200 & 15 &0.02  \\
  \end{tabular}\right\}$ $ \kappa + \Sigma_{\rm a}$
\end{tabular}  
\\
\hline
\begin{tabular}{l}
$\kern-\nulldelimiterspace\left.
  \begin{tabular}{p{3.2cm}p{1.5cm}p{1.5cm}p{1.5cm}p{1.5cm}}
\hspace{-0.7cm}\simname{MS\_Tc\_Z0.02}   & 1.152   & 100 & 25 &0.02  \\
  \end{tabular}\right\}$ $ \kappa  + T_{\rm core}$
\end{tabular}  
\\
\hline
\hline

\begin{tabular}{l}
$\kern-\nulldelimiterspace\left.
  \begin{tabular}{p{3.2cm}p{1.5cm}p{1.5cm}p{1.5cm}p{1.5cm}}
\hspace{-0.7cm}\simname{ML\_fid}   & 0.576   & 100 & 15 & 1.0  \\ 
\end{tabular}\right\}$ Fiducial lower mass model
\\ 
\end{tabular}  
\\ \hline
\begin{tabular}{l}
$\kern-\nulldelimiterspace\left.
  \begin{tabular}{p{3.2cm}p{1.5cm}p{1.5cm}p{1.5cm}p{1.5cm}}
\hspace{-0.7cm}\simname{ML\_Z0.1}   & 0.576   & 100 & 15 &0.1  \\
  \end{tabular}\right\}$ $ \kappa $ only
\end{tabular}  
\\
\hline
\begin{tabular}{l}
$\kern-\nulldelimiterspace\left.
  \begin{tabular}{p{3.2cm}p{1.5cm}p{1.5cm}p{1.5cm}p{1.5cm}}
\hspace{-0.7cm}\simname{ML\_Tc\_Z0.02}   & 0.576   & 100 & 25 &0.02  \\
  \end{tabular}\right\}$ $ \kappa  + T_{\rm core}$
\end{tabular}  
\\
\hline
\end{tabular}\\
\end{table*}

The treatment of the inner boundary conditions of the numerical simulations of PPDs needs special attention. 
A complication may arise if the inner boundary allows for matter to flow only in one direction, i.e., outflow only boundary condition from the disk to the sink cell.
The wave-like motions near the inner boundary result in a disproportionate flow through the sink-disk interface and it causes an artificial drop in the gas density near the inner boundary. 
We consider a carefully implemented inflow-outflow inner boundary condition for our simulations \citep{Vorobyov2018}.
Here, the material that passes to the sink cell through the sink-disk interface is redistributed between the central protostar and the sink cell.
Depending on the mass surface density and velocity gradients at the boundary, the material is allowed to flow from the sink cell back onto the active disk.
This method ensures that the innermost parts of the disk are unaffected by the proximity of the boundary and related numerical artefacts.
For all simulations presented in this study, the inner boundary is placed at 0.42 au. 
This allows us to evolve the innermost parts of the disk with an accuracy sufficient for capturing the sub-au scale behavior of the PPD.

For this study of PPDs in low metallicity environment, we conducted a total of nine hydrodynamic simulations as listed in Table~\ref{table:sims}.
We considered two masses of the parental cloud core -- 1.152 and 0.576 $M_\odot$ -- which will eventually form a solar-type dwarf star and a fully convective low mass star, respectively.
The prescripts ``\simname{MS}" and ``\simname{ML}" describe the final stellar mass.  
The fiducial models with a postscript ``\simname{fid}" are the standard models, conducted with a composition of solar metallicity against which the low metallicity results are compared.
We consider two values of low metallicities.
The postscript ``\simname{Z0.1}" corresponds to models with 10\% of the solar metallicity, such as the star-forming regions observed in the outer Galaxy.
The extremely low metallicity of 2\% of the solar value (``\simname{Z0.02}" models) may correspond to the young galaxies undergoing their first major episodes of star formation.
In accordance with previous numerical studies \citep[e.g.,][]{Boss02,Cai06}, we set sub-solar metallicity by scaling down the gas and dust opacities of the solar-metallicity disk by the corresponding factor.
Although the thickness of the MRI active surface layers as determined by the Galactic cosmic rays is uniform at about $100$~g~cm$^{-2}$ \citep{UN1981}, the lower dust content in the low metallicity PPDs may offer reduced rates of recombination on grain surfaces resulting in a larger penetration depth \citep{Yasui09}.
We thus demonstrate the effects of the increased MRI efficiency with two simulations (denoted by ``\simname{$\Sigma$a}" in the model name) with a higher value of $200$~g~cm$^{-2}$.
In the case of lowest metallicity environment of about $Z = 1\% Z_\odot$, the dominant cooling mechanism of dust continuum emission is much less effective during the initial stages of the core collapse phase \citep{Omukai05}.
This results in a higher background gas temperature as compared to the higher metallicity counterparts and higher disk infall velocities during the pre- and protostellar stages \citep{Vorobyov20}.
To investigate the effects of this phenomenon, we conducted two additional simulations with an increased initial gas temperature of the prestellar cloud core of 25 K, which was also set to the temperature of the background radiation (indicated by ``\simname{Tc}" in the model name). 
The default value of these temperatures was set to 15 K for rest of the simulations.
The last column of Table \ref{table:sims} summarizes the included physical effects in each model with a self-explanatory notation.

The simulations were conducted for 0.7 Myr and at the end the system may be considered to be in the T~Tauri phase where the envelope has already accreted onto the star-disk system and most of the burst activity is over.
Here, the time is measured from the start of the gravitational collapse of the core, i.e., the beginning of a simulation.
The radial and azimuthal resolution of the computational grid for the simulations was set to $256 \times 256$, for all simulations, with a logarithmic spacing in the radial direction and uniform spacing in the azimuthal direction.
The maximum grid resolution in the radial direction was $1.6 \times 10^{-2}$ au at the innermost cell of the disk ($0.417$ au), while the poorest resolution near the outer boundary of the cloud ($10210$ au for \simname{MS} simulations) was about $400$ au.
Note that any two simulations conducted with the same initial conditions and parameters do not produce identical disk evolution because of gravitationally unstable disks, however, the simulation results are always congruent and we have tested their convergence with the help of higher resolution models.

\section{Results}
\label{sec:results}
In this section we will present our results of disks formed in low metallicity environments.
In Section \ref{sec:largescale} we show the differences in disk structure and evolution on a global scale.
In Section \ref{subsec:opacity} the effects of opacity only are considered, while in Section \ref{subsec:opacity} we elaborate on added effects of increased active layer thickness and increased cloud core temperature.
In Section \ref{subsec:ML} we consider low mass cloud cores and finally in Section \ref{subsec:burst} an individual MRI burst is characterized.
Table \ref{table:results} summarizes some of the key results from all nine simulations for comparison, which will be explained in the upcoming sections.

\subsection{Large Scale Evolution}
\label{sec:largescale}

\begin{table*}
\begin{center}
\caption{Summary of Simulation Results}
\label{table:results}
\begin{tabular}{|lccccc|}
\hline
\hspace{0.5cm} Model Name \hspace{0.5cm} &  ${ M_{ \rm \ast, F} (M_{\odot}) }^{\star}$  &  ${ M_{ \rm disk, F} (M_{\odot}) }$  &  ${ M_{ \rm disk < 10 au, F} (M_{\odot}) }$ &   Embedded phase (Myr) &   Burst phase (Myr) \\
\hline
\simname{MS\_fid}   &   0.816   & 0.227 & 0.015 & 0.242 & 0.552  \\ 
\simname{MS\_Z0.1}   & 0.772   & 0.282 & 0.078 & 0.242 & 0.414 \\ 
\simname{MS\_Z0.02}   & 0.782   & 0.285 & 0.084 & 0.242 & 0.274  \\
\hline
\simname{MS\_$\Sigma$a\_Z0.1}   & 0.821   & 0.220 & 0.011 & 0.242 & 0.415 \\ 
\simname{MS\_$\Sigma$a\_Z0.02}   & 0.804   & 0.236 & 0.048 &0.243 &0.220 \\
\simname{MS\_Tc\_Z0.02}   & 0.840   & 0.187 & 0.082 & 0.113 &0.129 \\
\hline
\simname{ML\_fid}   & 0.459   & 0.042 & $2.34 \times 10^{-3}$ & 0.095 & 0.25 \\ 
\simname{ML\_Z0.1}   & 0.458   & 0.044 & $3.38 \times 10^{-3}$  &0.096 & 0.18 \\
\simname{ML\_Tc\_Z0.02}   & 0.478   & 0.022 & $6.05 \times 10^{-3}$ & 0.044 & -- -- \\
\hline
\end{tabular}  
\end{center}
$^{\star}$ The subscript ``F" denotes final values measured at 0.7 Myr.
\end{table*}

\begin{figure*}
\centering
  \includegraphics[width=18cm]{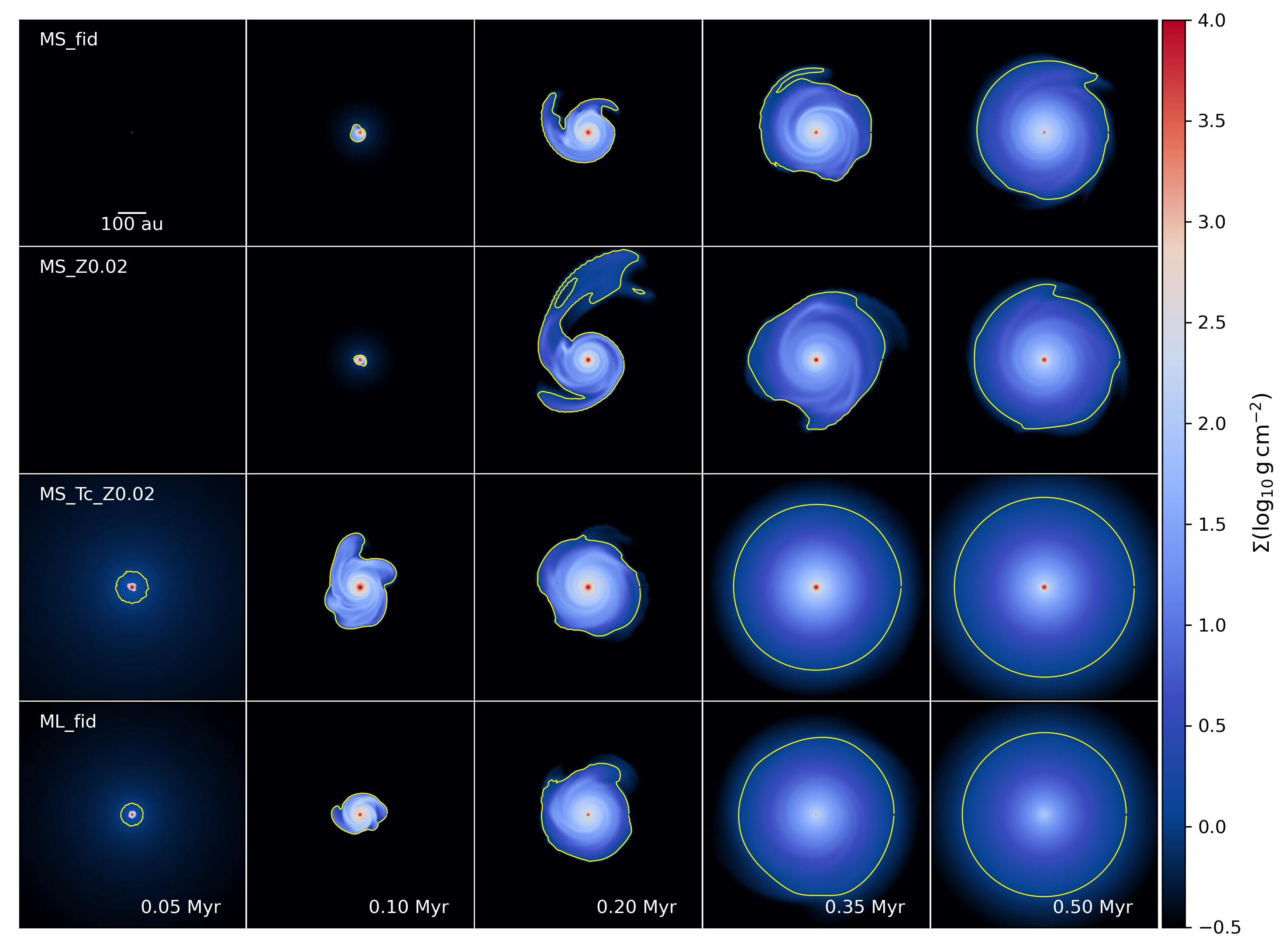}
\caption{Evolution of the disk gas surface density distribution (in the units of log$_{10}$ g~cm$^{-2}$) for the {four models} -- \simname{MS\_fid}, {\simname{MS\_Z0.02}}, \simname{MS\_Tc\_Z0.02} and \simname{ML\_fid} -- over a region of 500 x 500 au, showing the large-scale disk structure. The yellow contours mark $\Sigma =1$ g\,cm$^{-2}$ levels, indicating approximate outer boundary of the disks. }
\label{fig:global}
\end{figure*}

As we shall see later, a lower metallicity has a significant effect on innermost parts of a PPD.
However, in certain cases the large-scale structures and global evolution of the disk are also noticeably affected.
Figure \ref{fig:global} shows evolution of the disk in {four simulations -- \simname{MS\_fid}, \simname{MS\_Z0.02},} \simname{MS\_Tc\_Z0.02} and \simname{ML\_fid} -- over a region of $500 \times 500$ au$^2$ during the first 0.5 Myr.
Note that the time duration between the snapshots is not uniform.
The yellow contours mark $\Sigma =1$ g\,cm$^{-2}$ level indicating the approximate extent of the disk.
Consider the solar mass, solar metallicity model, \simname{MS\_fid}, in the top row.
The disk was formed around 0.08 Myr (therefore not observed in the first frame) and then showed viscous spread as it evolved.
The initial phase was dominated by vigorous GI instability and associated large-scale spirals.
The initial ratio of kinetic to gravitational energy was chosen low enough so that the disks will typically form no clumps.
As seen in the last panel at 0.5 Myr the disk eventually became weakly unstable and progressively more axisymmetric. 
This approximate symmetry was maintained till the end of the simulation.
Evolution on this large scale for models \simname{MS\_Z0.1}, \simname{MS\_Z0.02}, \simname{MS\_$\Sigma$a\_Z0.1} and \simname{MS\_$\Sigma$a\_Z0.02} was very similar to \simname{MS\_fid} model with two key differences.
Even at this scale, the innermost regions of $\lessapprox  10$ au showed a consistently larger gas surface density (e.g., similar to \simname{MS\_Tc\_Z0.02} simulation in the last snapshot).
The lower metallicity models were also prone to more GI activity and associated spirals, although no fragmentation was observed.
{The second row of Figure \ref{fig:global} shows the evolution for model \simname{MS\_Z0.02} where both of these differences can be noticed.}
The metal-poor environment offers a lower opacity and hence more efficient cooling of the disk, which in turn can favor GI. 
See \cite{Vorobyov20} for a detailed study of accretion bursts caused by GI fragments in low metallicity PPDs.

Amongst the solar mass simulations, the most significant difference on the large scale disk evolution was observed for \simname{MS\_Tc\_Z0.02} simulation.
The increased cloud core temperature (25 K as opposed 15 K for the other \simname{MS} models) affects the thermal evolution of the disk resulting larger infall velocities \citep{Vorobyov20}.
As a consequence, the mass infall rate on the disk was higher in the early stages and the disk showed a faster growth as well as a shorter period of GI activity.
As seen in the second row of Figure \ref{fig:global}, the disk was already formed at 0.05 Myr and appeared more developed at 0.35 Myr than \simname{MS\_fid} at 0.5 Myr.
We will elaborate on this accelerated evolution of \simname{MS\_Tc\_Z0.02} in Section \ref{subsec:sigmaTc}.
The last row of Figure \ref{fig:global} shows the evolution of surface density for the lower mass solar metallicity model \simname{ML\_fid}.
Disk evolution for a lower mass cloud core occurs on a faster timescale because of the limited mass reservoir in the parental cloud core. 
The disk showed some GI activity and associated spirals in the initial stages, but no clumps were formed in any of the low mas models.
The extent of the disk was smaller at later times and showed less accumulation of gas in the innermost regions.
The trends for the rest of the lower metallicity simulations were similar to the higher mass models. The model \simname{MS\_Z0.1} showed marginal differences at a large scale in the direction of more vigorous GI, while \simname{MS\_Tc\_Z0.02} showed an accelerate evolution.
In metal poor environments, most notable effects occur at smaller scales at less than 10 au of the inner disk, which we will focus on next.

\subsection{Effects of Opacity}
\label{subsec:opacity}

\begin{figure*}[t]
\hspace{3.3cm} \simname{MS\_fid}  \hspace{4cm} \simname{MS\_Z0.1}  \hspace{3.5cm} \simname{MS\_Z0.02}  \\
\begin{tabular}{l}
\vspace{-0.27cm}\includegraphics[width=0.992\textwidth]{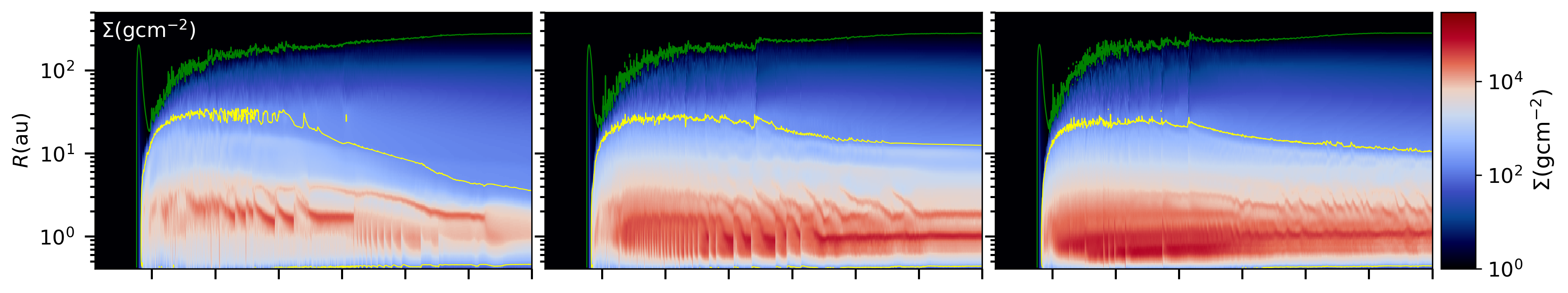} \\
\vspace{-0.27cm}\includegraphics[width=\textwidth]{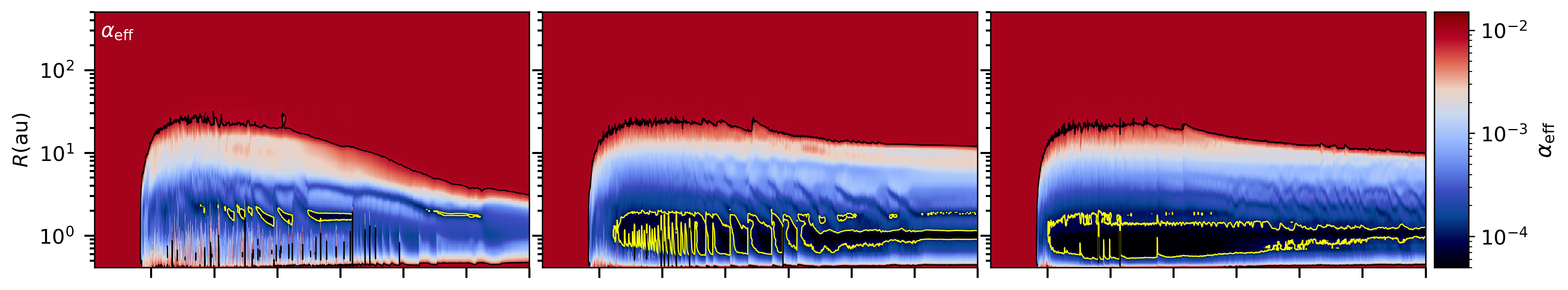} \\
\vspace{-0.27cm}\includegraphics[width=0.995\textwidth]{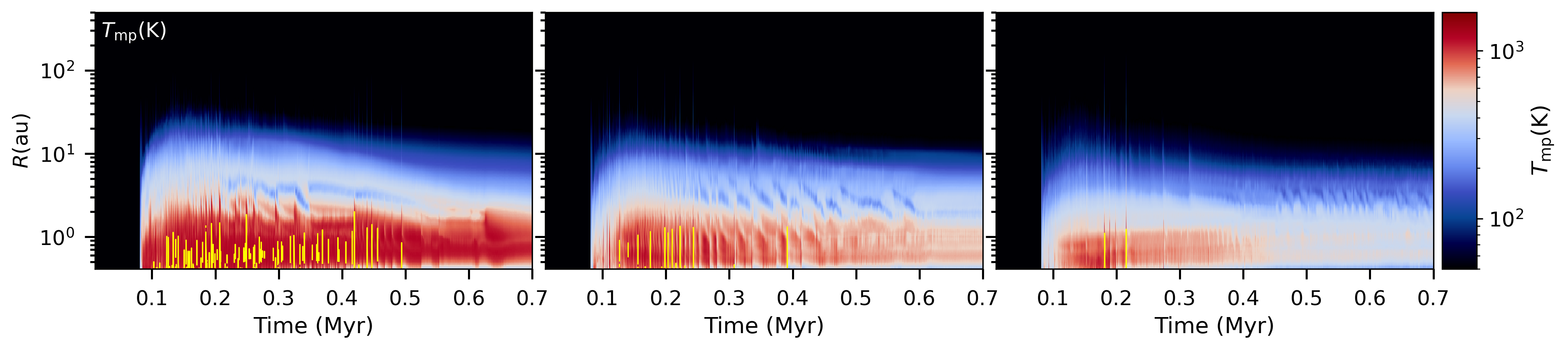} \\  
\end{tabular}
\caption{
The spacetime plots for the three models -- \simname{MS\_fid}, \simname{MS\_Z0.1} and \simname{MS\_Z0.02} -- are compared. 
The rows depict the evolution of azimuthally averaged quantities -- $\Sigma$, $\alpha_{\rm eff}$ and $T_{\rm mp}$. 
The green and yellow curves in $\Sigma$ show 1 and 200 ${\rm g~cm^{-2}}$ contours respectively.
The black contour in $\alpha_{\rm eff}$ shows the extent of the dead zone,
while the yellow lines shows $10^{-4}$ contour. 
The yellow lines in the temperature plots show $T_{\rm crit}=1300$ K contours. 
}
\label{fig:spacetimeN1}
\end{figure*}

In this section we elaborate on general features of the fiducial solar metallicity simulation \simname{MS\_fid}. We compare this to the low metallicity simulations \simname{MS\_Z0.1} and \simname{MS\_0.02} with $Z=0.1 Z_\odot$ and $Z=0.02 Z_\odot$, respectively. Note that in the latter models, metal poor environment is emulated by scaling down the gas and dust opacities by the same fraction.   
In Figure \ref{fig:spacetimeN1} the evolution of the inner disk is shown with the help of spacetime diagrams.
The ordinate extends from the inner radius of the computational domain of 0.42 au to 500 au on a logarithmic scale, while the abscissa is time in Myr.
The first column represents simulation \simname{MS\_fid} with solar metallicity and the next two columns show models \simname{MS\_Z0.1} and {\simname{MS\_Z0.02}}, respectively.
The rows compare the azimuthally averaged profiles of quantities -- gas mass surface density, effective $\alpha$ and the midplane temperature. 

Consider the first column for the simulation \simname{MS\_fid}. 
In the first row, the green contour plotted at $\Sigma=1$ g~cm$^{-2}$ shows the approximate extent of the disk. 
The yellow contour marks ${2 \times \Sigma_{\rm a}=200}$ g~cm$^{-2}$ level, incorporating both sides of the disk and thus, the region outside this boundary is fully MRI active. 
The disk formed at about 0.08 Myr and axisymmetric ring structures soon began appearing in the gas surface density.
In \cite{Kadam19} we have described the evolution of the rings in detail, which form at a distance of a few au due to viscous torques occurring at the inner boundary of the dead zone.
The simulation \simname{model1\_T1300\_S100} in \cite{Kadam19} was identical to \simname{MS\_fid}, but conducted with twice the spatial resolution.
The salient features of \simname{MS\_fid} remain consistent with the earlier results, indicating convergence of the numerical simulations.
{We will demonstrate the congruency across spatial resolutions in detail by comparative analysis of these two simulations in Appendix \ref{appendixA}.} 
The second row of Figure \ref{fig:spacetimeN1} compares ${\rm \alpha_{eff}}$, which has a constant value of 0.01 for the fully MRI active disk in the outer regions.
The black contour shows the approximate extent of the dead zone, defined as the region where the $\alpha_{\rm eff}$ is below 80\% of this maximum value.  
The yellow contours mark $10^{-4}$ level, thus indicating the regions of low viscosity which coincide with the gaseous rings.
The third row shows the midplane temperature with yellow contours marking the MRI activation temperature, $T_{\rm crit}=1300 K$.
The midplane temperature in the innermost regions frequently exceeded this value for a short time and some of these temperature increases were associated with MRI outbursts.

When comparing the effects of lower metallicity on the disk structure in Figure \ref{fig:spacetimeN1}, regions outside of about 10 au were very similar for all three simulations.
The green contour in $\Sigma$ indicating the disk extent, and the yellow contour in the same plots marking the fully MRI active outer disk, showed almost identical evolution. 
However, the inner disk structure was markedly different. 
At a lower metallicity clear formation of successive rings was not observed; instead, a progressively broader region of gas accumulation was formed.
The surface density of the accumulated gas increased as the metallicity decreased.
The $\alpha_{\rm eff}$ essentially varies inversely to total $\Sigma$ (Equation \ref{eq:alpha}), hence the lower metallicity resulted in a deeper dead zone in terms of disk viscosity.
This is clearly observed with the $\alpha_{\rm eff}=10^{-4}$ contours in the second row.
The dead zone also became broader and moved inward, in general, at low metallicities.
With a decrement in metallicity, the midplane temperature of \simname{MS\_Z0.1} and \simname{MS\_Z0.02} tended to be proportionally cooler as compared to the fiducial model.
The opacity was assumed to be proportional to the dust content and hence the metallicity of the disk.
With a lower opacity, the midplane of the disk can cool more efficiently, lowering the midplane temperature.
The disk viscosity is not only proportional to $\alpha_{\rm eff}$, but also to the local sound speed.
With a decreasing temperature, the sound speed also decreased, resulting in further decrement in viscosity.
This lower viscosity formed a narrower bottleneck to the mass transport, thus contributing to the observed trends with a lower metallicity -- accumulation of more gas as well as lower $\alpha_{\rm eff}$ in the inner disk region.

\begin{figure*}[t]
\hspace{3.6cm} \simname{MS\_fid}  \hspace{4.5cm} \simname{MS\_Z0.1}  \hspace{4cm} \simname{MS\_Z0.02}  \\
\begin{tabular}{r}
\vspace{-0.273cm}\includegraphics[width=\textwidth]{compMdot_finalN1.png} \\
\vspace{-0.26cm}\includegraphics[width=0.988\textwidth]{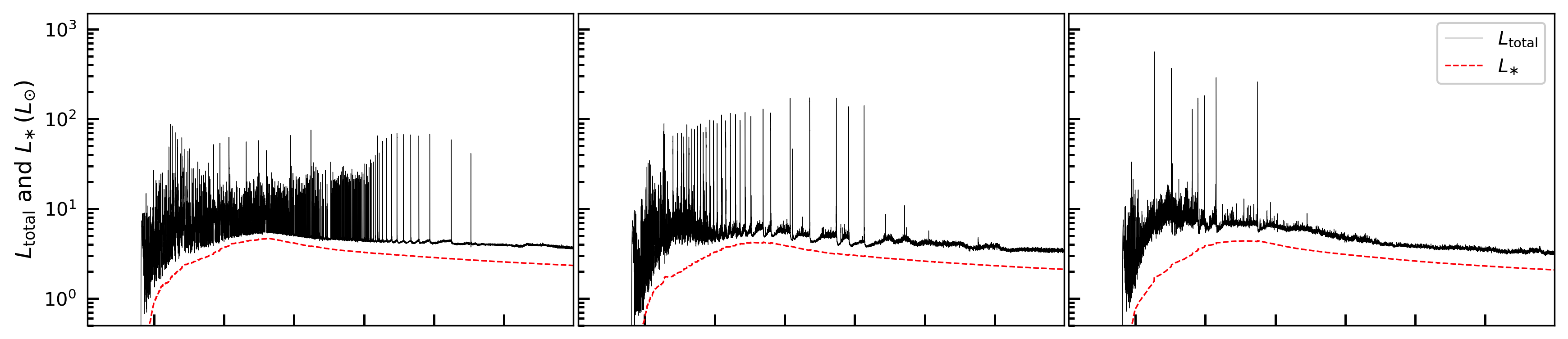} \\
\vspace{-0.25cm}\includegraphics[width=0.996\textwidth]{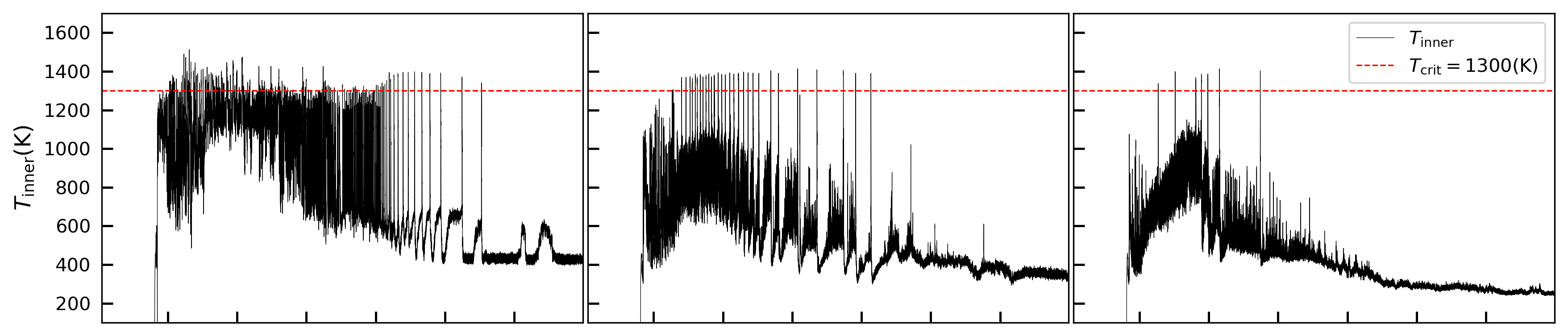} \\  
\includegraphics[width=0.984\textwidth]{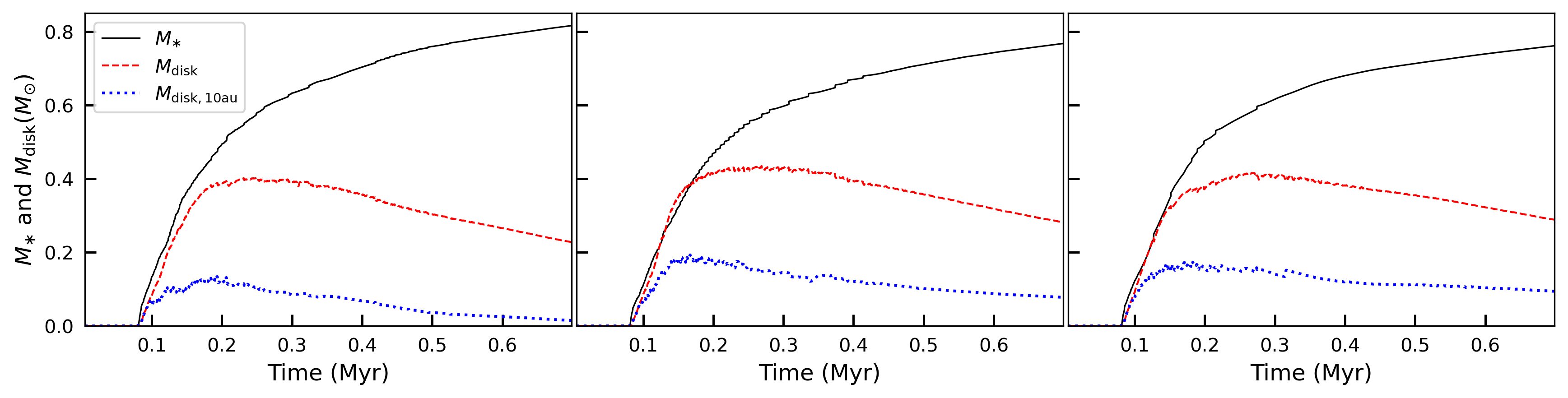} \\
\end{tabular}
\caption{The evolution of time-dependent quantities is compared for the three models -- \simname{MS\_fid}, \simname{MS\_Z0.1} and \simname{MS\_Z0.02} 
-- during the entire burst-phase.
The rows show total and stellar luminosities, mean mass accretion rate and cloud core infall rate, temperature at the inner computational boundary, as well as stellar and disk masses (total and that of the inner disk at $< 10$ au), respectively.
The three boxes in the first row highlight the three modes of accretion rate variability encountered in the simulations. Box A corresponds to excursion of the inner edge of the dead zone across the computational boundary, Box B marks the MRI bursts and Box C corresponds to variability due to vigorous GI activity. The arrow marks the particular MRI burst investigated in detail in Section \ref{subsec:burst}
}
\label{fig:1dN1}
\end{figure*}

In Figure \ref{fig:1dN1} we compare the outbursting activity of the same three simulations -- \simname{MS\_fid}, \simname{MS\_Z0.1} and \simname{MS\_Z0.02} -- with respect to some of the time dependent quantities. 
The first row compares the accretion rate onto the central star and infall rate on the disk from the cloud core ($\dot{M_\ast}$ and $\dot{M}_{\rm infall}$, respectively).
The three boxes (A, B and C) highlight the three modes of accretion variability that will be discussed later in this section.
The $\dot{M_\ast}$ is calculated as mass passing through the sink cell of the inflow-outflow inner boundary, with the ratio of mass accreting onto the star to that going into the sink cell set to 95\%:5\% \citep{Kadam19}.
Our model also assumes that 10\% of the mass that passes through the sink cell is lost via jets or outflows before this partitioning.
$\dot{M_{\rm infall}}$ is considered to be the mass flux at a radial distance of 1000 au from the center. 
The second row shows the total and stellar photospheric luminosities, $L_{\rm total}$ and $L_\ast$, respectively.
The total luminosity is the sum of accretion and the stellar photospheric luminosity, where the latter quantity is calculated through the precomputed tracks of \cite{DAntona97}.
The third row compares the midplane temperature at the inner boundary of the disk ($T_{\rm inner}$).
The last row shows the evolution of the stellar and the the disk mass ($M_\ast$ and $M_{\rm disk}$, respectively), as well as the mass of the inner disk less than 10 au ($M_{\rm disk, 10 au}$).
In this study, we define end of the ``burst phase" of a PPD as the time up to the last major outburst observed in the simulation.
We consider the end of the ``embedded phase" of the disk as the time instance when the infalling envelope retains less than 10\% of the initial core mass.
Table \ref{table:results} lists the final masses and duration of the above two phases at the end of all simulations.
Although the dead zone may continue to exist, it is unlikely that the systems will show additional large amplitude, FUor-like outbursts after the termination of the simulations at 0.7 Myr, i.e., after the initial burst phase is over \citep{Zhu10b, Bae14}.

Consider the first row of Figure \ref{fig:1dN1}. 
The overall, qualitative behavior of the mass accretion history is consistent with previous 1D and 2D simulations \citep{Zhu10b, Bae14,VB2015}.
The outbursts were superimposed on a steadily decreasing background accretion rate which stabilized at a few times $10^{-7} M_{\odot} {\rm yr}^{-1}$ at later times.  
The early burst phase was characterized by high accretion variability reflecting the dynamical nature of the disk, primarily the presence of GI induced spiral waves.
As the mass infall rate decremented with time, the mass-loading of the disk diminished and vigorous GI activity could not be sustained, and the accretion variability diminished. 
This variability is best observed in \simname{MS\_Z0.02} simulation, and highlighted by Box C in the figure.
 {Typical FU Orionis-like eruptions are characterized by a sudden increase in $\dot{M_\ast}$ by more than an order of magnitude.
These large amplitude eruptions are MRI-outbursts related to sudden triggering of MRI in the inner disk \citep{Kadam20}.}
This mode of accretion variability caused by MRI outbursts is highlighted by Box B.
The arrow marks the individual outburst that will be investigated in detail in the context of metal-poor environment in Section \ref{subsec:burst}.
The MRI bursts were clearly distinguishable from the GI related variability.
The outbursts were initially clustered together, superimposed on the accretion variability. The duration between two outbursts increased with time, eventually leading to termination of the burst phase.
{The simultaneous presence of both MRI and GI in the disk raises the possibility of recently proposed ``spiral wave dynamo", especially in the early times \citep{RL18, RL19, Deng20}.
The GI spiral waves can initiate a magnetohydrodynamic dynamo in the disk that can amplify initial magnetic field. 
The associated magnetic torques can drive mass accretion and may revive the dead zone even before MRI outburst is triggered. 
The action of such non-ideal magnetohydrodynamic effects should be considered in future investigations.}

When considering how the lower metallicity affects accretion rates, two trends are immediately apparent.
First, the duration of the burst phase shortened with a decrease in the metal content.
The burst phase ended at approximately 0.55, 0.42 and 0.27 Myr for \simname{MS\_fid}, \simname{MS\_Z0.1} and \simname{MS\_Z0.02}, respectively.
With the duration of the embedded phase remaining constant at about 0.24 Myr, most of the outbursting activity in the lower metallicity models was hidden from a direct view.
Second observation is that the amplitude individual bursts increased with decreasing metallicity.
This can also be observed in the luminosity plots in the second row. 
All of the bursts in \simname{MS\_fid} simulation were less than 100 $M_\odot$ in the luminosity output, and became progressively more powerful, typically above 100 $M_\odot$ for the lower metallicity counterparts.
We conjecture that the reason behind the shortening of the burst phase with decreasing metal content was the overall decrease in temperature of the innermost parts of the disk.
With the inclusion of dead zone viscosity, the MRI burst is triggered by the viscous heating near the inner edge of the dead zone and propagates outward in an inside-out manner (see Section \ref{subsec:burst}).
As seen in Figure \ref{fig:spacetimeN1}, the metal poor disks cool very efficiently which can inhibit such triggering of the outbursts.
In the same figure, the spacetime plot of the gas mass surface density shows accumulation of larger amount of mass at lower metallicities.
Thus making more amount of material available for accretion during an outburst, explaining the trend in the intensity of the bursts.

The third row of Figure \ref{fig:1dN1} shows the temperature at the inner boundary of the disk.
For the lower metallicity models, the luminosity bursts coincided with the occasional crossings of $T_{\rm inner}$ above the critical temperature of 1300 K.
This is expected for an MRI outburst as the underlying mechanism depends on temperature-dependent viscosity.
For \simname{MS\_fid} simulation, however, the temperature at the inner boundary constantly fluctuated across this critical threshold.
Here we mention one caveat of our model. 
The computational domain of the simulations encompasses region outside of 0.42 au and despite carefully constructed inflow-outflow boundary, the simulation results reflect the conditions at this location in the disk.
The real mass accretion rate onto the star may be modified by the physical conditions and mechanisms that may operate inside this innermost disk regions, such as magnetospheric accretion in low mass YSOs.
In the case of layered disks such as modelled here, as we move closer to the central star, the disk changes its state from having a dead zone to a fully MRI active region due to thermal ionization of the gas.
Part of the fluctuations that are observed in $T_{\rm inner}$, and consequently the accretion rate onto the star, are due to the radial excursion of the inner boundary of the dead zone across the inner boundary of the computational domain. 
On the one hand, when the boundary is placed further out, important phenomena occurring in the inner few au such as MRI outbursts are not captured at all.
On the other hand, additional complications may arise when the boundary is placed sufficiently close to the star.
The latter case seems to hold true for \simname{MS\_fid}, where it is difficult to differentiate between real accretion activity and movement of the inner fully MRI-active zone.
This third mode of irregularity in the mass accretion rate is highlighted by Box A in Figure \ref{fig:1dN1}.
The magnitude of this accretion variability was typically few times smaller than the MRI outbursts.
The inner edge of the dead zone moved inward and away from the computational boundary for the metal poor disks due to the lower temperatures.
Thus, this anomalous variability was diminished for the lower-metallicity models.
This remains a general drawback of disk simulations which do not model the disk reaching all the way to the stellar surface or magnetosphere.
The overall results of this study are not affected by these fluctuations, as our results remain consistent with the previous investigations.

The last row of Figure \ref{fig:1dN1} compares the evolution of the stellar and disk masses.
For all three simulations, the disk remained substantially massive during the burst phase and was a large fraction of the stellar mass at the end of the simulations at 0.7 Myr.
With a lower metallicity, the final stellar mass marginally decreased, while the total disk mass was marginally larger (see Table \ref{table:results} for numerical values).
Although the total final disk masses were comparable, the innermost parts were substantially massive at low metallicities.
At the end of the simulations, only 6\% of the total disk mass in \simname{MS\_fid} was contained within the inner 10 au region, while about 30\% of the disk mass was in the inner 10 au for the lowest metallicity \simname{MS\_Z0.02} model. 
This indicates that the dead zone in a low metallicity environment formed a narrower bottleneck to the mass transport, affecting the long-term evolution of the system. 
The observations suggest that frequency of disk harboring stars (i.e., disk fraction) decreases significantly at low metallicity, indicating a shorter disk lifetime \citep{Yasui10}.
Our finding of more massive inner disks in metal poor environments suggests that the disk dispersal via photoevaporation occurring at late stages of these disks may be more efficient than previously estimated \citep[e.g.,][]{ErcolanoClarke10}. 
The frequency of close-in super-Earths is observed to be almost independent of the host star's metallicity \citep{Petigura18}.
The massive inner disk may facilitate dust accumulation, growth and formation of planetesimals, explaining this trend.
The post-burst phase accretion rates were similar across the three models near the end of simulations.
The transport of accumulated mass in the inner disk at later times may be responsible for observed trends of higher accretion rates at low metallicities \citep{Spezzi12,DeMarchi13}.

\subsection{Active layer thickness and cloud core temperature}
\label{subsec:sigmaTc}

For a magnetically layered PPD, the MRI turbulence depends on ionization degree of the disk gas, which essentially determined by the details of the ionization and recombination rates. 
Galactic cosmic rays are considered to be the major source of ionization at the distance of a few au from the central star, which ionize a nearly uniform gas surface density of about $96$~g~cm$^{-2}$ \citep{UN1981}.  
However, there is considerable uncertainty as strong winds of a protostar or accretion outflows can shield the inner disk \citep{Cleeves2013}.
At median T Tauri luminosities, the combined effect of stellar FUV photons and X--rays is at least an order of magnitude smaller \citep{Bergin2007, Cleeves2013}.
Thus a canonical value of $100$~g~cm$^{-2}$ is used for the thickness of the MRI active surface layer in our models presented in Section \ref{subsec:opacity}.
In a low metallicity environment, we can expect a proportionally lower dust content to be present in the disk.
As the recombination on grain surfaces is reduced, this would result in a thicker active layer as compared to solar metallicity disks \citep{Fromang02}. 
However, a low metallicity also results in fewer molecular ions in the disk, which contribute towards a reduced ionization fraction \citep{Sano00}.
It is speculated that in this tug of war, the former process should dominate and a thicker active layer would form \citep{Yasui09}.
The exact effects are difficult to calculate due to their dependence on long-term dust as well as chemical evolution of the disk and have not been published in the literature.
In this section, we assumed that the active layer thickness increases and approximately doubles in metal poor disks.
Although the magnitude is arbitrary, the comparison will give us an insight into how the increase in active layer thickness can affect the disk evolution.

Another effect considered in this section is the increase in the initial temperature of the molecular cloud core in a low metallicity environment. 
With global numerical simulations of PPD formation which incorporate a sophisticated heating and cooling treatment with separate gas and dust temperatures, \cite{Vorobyov20} showed that in the case of extreme metal poor disks ($Z=0.01 Z_\odot$) the gas and dust temperatures show significant decoupling during the core collapse phase.
This was because at low densities, the dust continuum emission is a dominant cooling mechanism \citep{Omukai05}.
The disks with extremely low metallicity cannot cool efficiently, leading to an increase in the gas temperature as well as increased infall velocities than their higher metallicity counterparts.
Thus for the lowest metallicity case, we also conducted simulations with an increased cloud core temperature of 25 K, as opposed to a canonical value of 15 K (see Table \ref{table:sims}).

\begin{figure*}[t]
\hspace{3cm} \simname{MS\_$\Sigma$a\_Z0.1}  \hspace{3cm} \simname{MS\_$\Sigma$a\_Z0.02}  \hspace{3cm} \simname{MS\_Tc\_Z0.02}  \\
\begin{tabular}{l}
\vspace{-0.25cm}\includegraphics[width=0.99\textwidth]{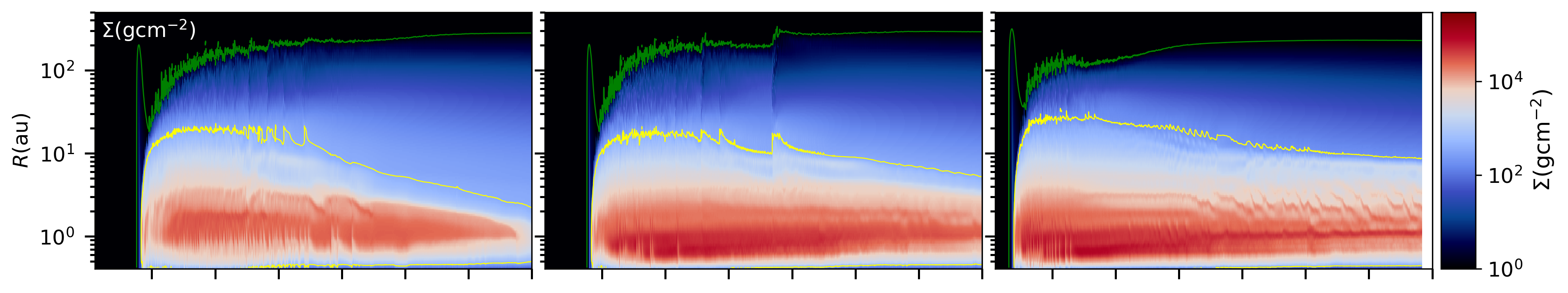} \\
\vspace{-0.25cm}\includegraphics[width=\textwidth]{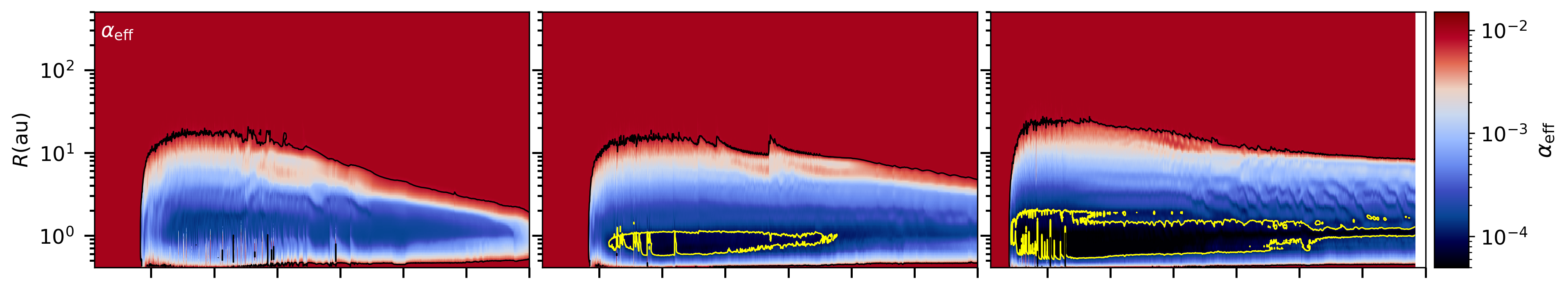} \\
\vspace{-0.25cm}\includegraphics[width=0.995\textwidth]{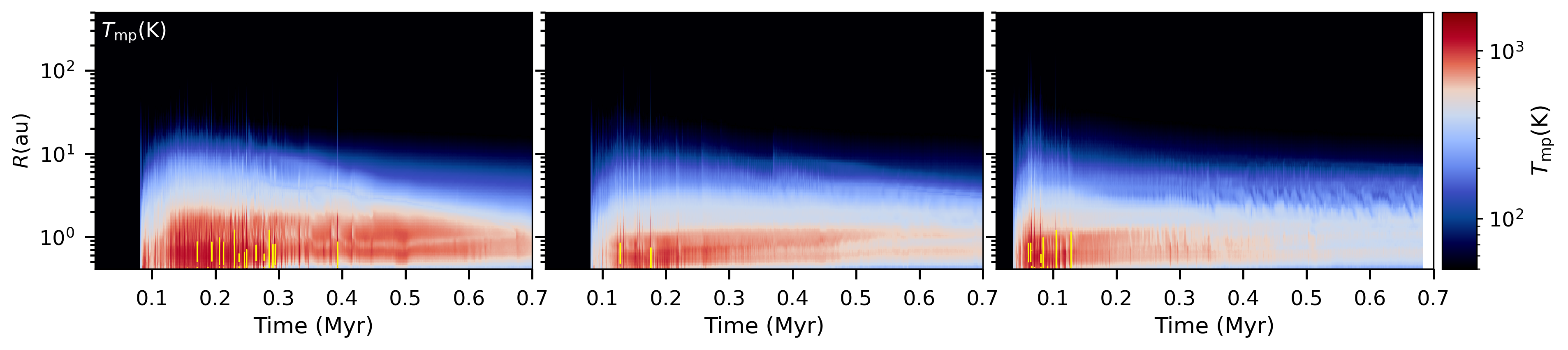} \\  
\end{tabular}
\caption{
The spacetime plots for the three models -- \simname{MS\_$\Sigma$a\_Z0.1}, \simname{MS\_$\Sigma$a\_Z0.02} and \simname{MS\_Tc\_Z0.02} -- are compared. 
The rows depict the evolution of azimuthally averaged quantities -- $\Sigma$, $\alpha_{\rm eff}$ and $T_{\rm mp}$.
The green and yellow curves in $\Sigma$ show 1 and $\Sigma=2 \times \Sigma_{\rm a}$ contours, respectively.
Note that the value of $\Sigma_{\rm a}$ is 200 ${\rm g~cm^{-2}}$ for the first two simulations and it equals 100 ${\rm g~cm^{-2}}$ otherwise.
The black contour in $\alpha_{\rm eff}$ shows the extent of the dead zone,
while the yellow lines shows $10^{-4}$ contour. 
The yellow lines in the temperature plots show $T_{\rm crit}=1300$ K contours. 
}
\label{fig:spacetimeN2}
\end{figure*}

In Figure \ref{fig:spacetimeN2} we present the low metallicity models with an increased active layer thickness -- \simname{MS\_$\Sigma$a\_Z0.1} and \simname{MS\_$\Sigma$a\_0.02} -- and \simname{MS\_Tc\_Z0.02} simulation which includes evolution with an increased cloud core temperature.
In order to demonstrate the effects on an increased $\Sigma_{\rm a}$ only, we compare the first two models with their canonical counterparts, \simname{MS\_Z0.1} and \simname{MS\_0.02}, presented in Section \ref{subsec:opacity}, respectively. 
Consider the first row of Figure \ref{fig:spacetimeN2} showing the spacetime diagram of gas surface density.
The yellow contours in these first two models show $2 \times \Sigma_{\rm a} = 200$~g~cm$^{-2}$ levels.
The plots show a lesser accumulation of gas in the vicinity of the dead zones for both \simname{MS\_$\Sigma$a\_Z0.1} and \simname{MS\_$\Sigma$a\_0.02} simulations.
When considering the spacetime diagrams of $\alpha_{\rm eff}$, \simname{MS\_$\Sigma$a\_Z0.1} did not show the yellow $10^{-4}$ contours, indicating a shallower depth of the dead zone.  
The radial extent of the dead zone was also smaller as compared to \simname{MS\_Z0.1}.
Similar trends were observed for \simname{MS\_$\Sigma$a\_0.02}, where the dead zone formed was smaller in terms of both radial and temporal extent.
With the doubling of the active layer, the dead zone formed a less restrictive bottleneck for angular momentum and mass transport.
From Equation \ref{eq:alpha}, the above trends in both surface density and $\alpha_{\rm eff}$ are as expected.

The third row of Figure \ref{fig:spacetimeN2} shows the midplane temperature, which was marginally increased in the innermost parts, i.e., in the vicinity of 1 au, for both \simname{MS\_$\Sigma$a\_Z0.1} and \simname{MS\_$\Sigma$a\_0.02}, as compared to the corresponding models with $\Sigma_{\rm a} = 100$~g~cm$^{-2}$.
This can be explained by a general increase in viscous heating caused by a thicker active layer.
This indicates a negligible effect of an increased $\Sigma_{\rm a}$ on the evolution of the outer disk.

We compare the third simulation \simname{MS\_Tc\_Z0.02} with \simname{MS\_Z0.02} for demonstrating the effects of increased core temperature.
In Figure \ref{fig:spacetimeN2}, the evolution of gas surface density, extent of the dead zone as measured by $\alpha_{\rm eff}$ and the inner disk temperatures look very similar for the two cases, with one key difference.
The disk was formed at an earlier time at approximately 0.02 Myr in \simname{MS\_Tc\_Z0.02}, as compared to the canonical counterpart where the disk was formed at 0.08 Myr. 
The mass infall rate of the cloud core is proportional to $3/2^{\rm th}$ power of the gas temperature \citep{Shu77}.
The higher gas temperature (10 K higher than the standard value of 15 K used for \simname{MS\_Z0.02}) in the inner parts of collapsing core caused a proportional increase in the mass infall rate.
The resulting accelerated evolution of the system manifested as an early formation of the disk.

\begin{figure*}[t]
\hspace{3cm} \simname{MS\_$\Sigma$a\_Z0.1}  \hspace{3.8cm} \simname{MS\_$\Sigma$a\_Z0.02}  \hspace{3.5cm} \simname{MS\_Tc\_Z0.02}  \\
\begin{tabular}{r}
\vspace{-0.25cm}\includegraphics[width=\textwidth]{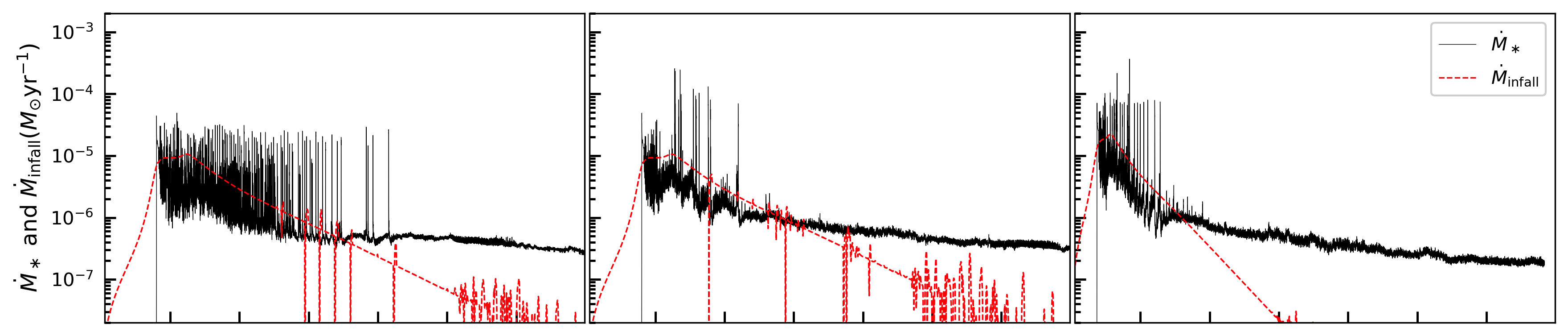} \\
\vspace{-0.25cm}\includegraphics[width=0.988\textwidth]{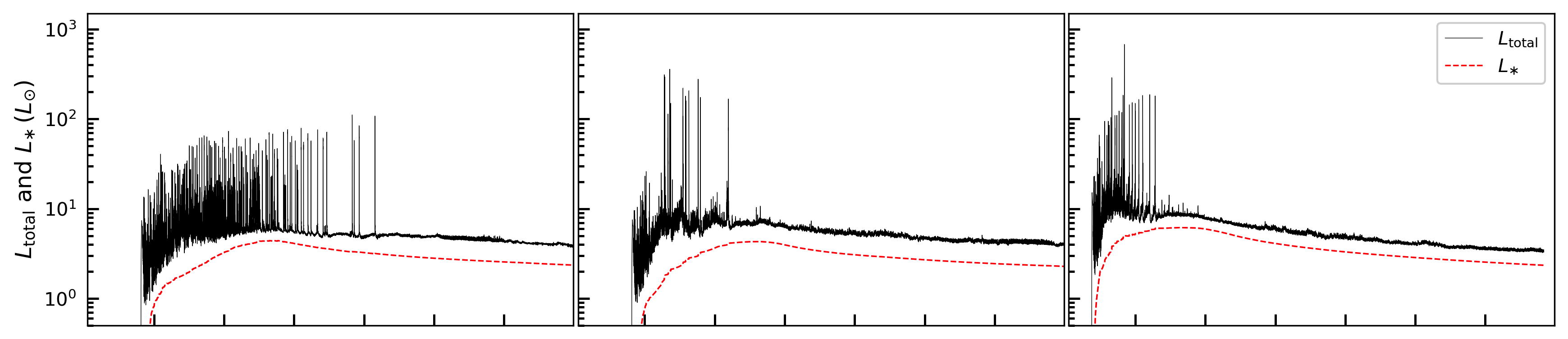} \\
\vspace{-0.25cm}\includegraphics[width=0.996\textwidth]{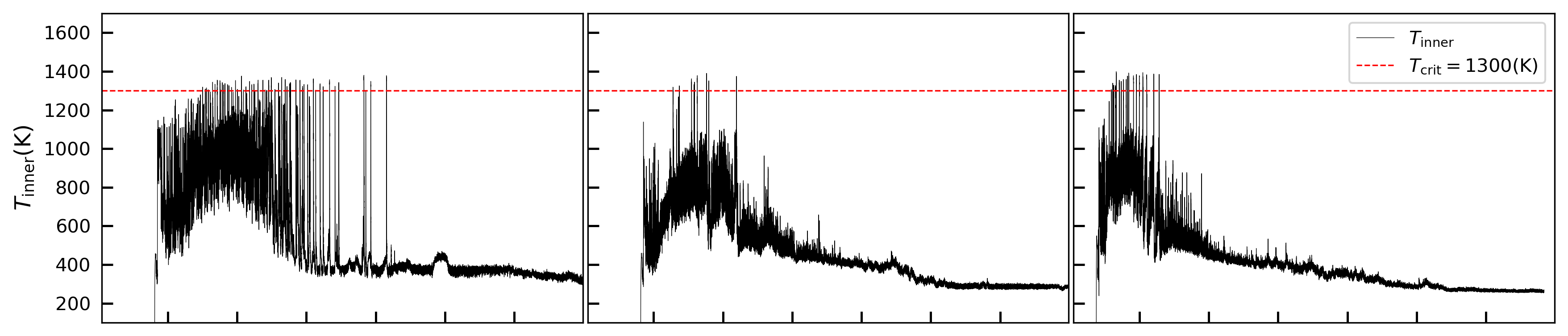} \\  
\includegraphics[width=0.984\textwidth]{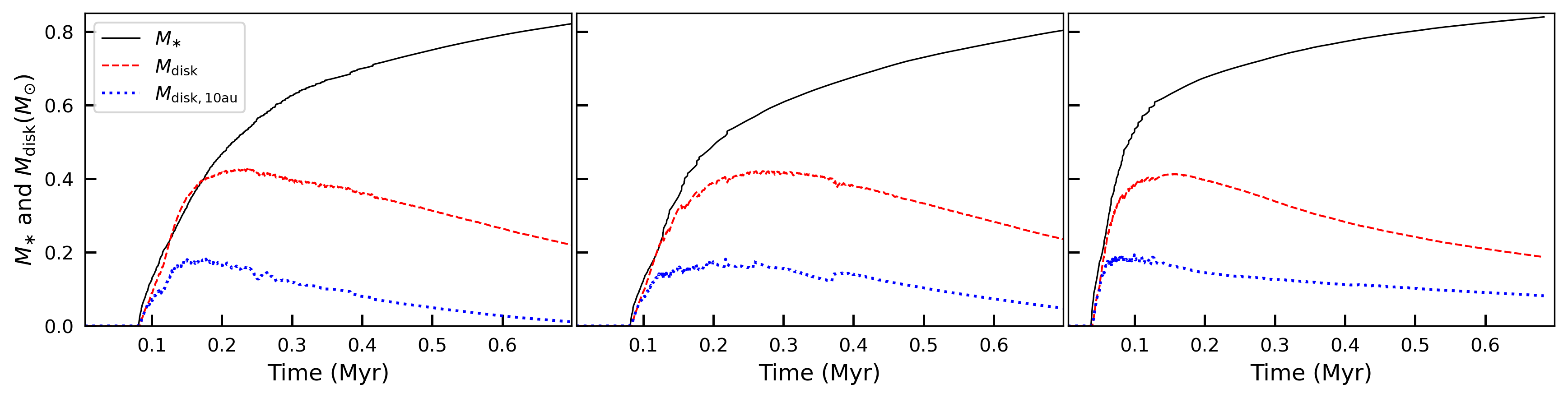} \\
\end{tabular}
\caption{The evolution of time-dependent quantities is compared for the three models -- \simname{MS\_$\Sigma$a\_Z0.1}, \simname{MS\_$\Sigma$a\_Z0.02} and \simname{MS\_Tc\_Z0.02} 
-- during the entire burst-phase.
The panels show total and stellar luminosities, mean mass accretion rate and cloud core infall rate, temperature at the inner computational boundary, as well as stellar and disk mass, respectively.
}
\label{fig:1dN2}
\end{figure*}

In Figure \ref{fig:1dN2} we present the quantities related to the episodic accretion of the same three models -- \simname{MS\_$\Sigma$a\_Z0.1}, \simname{MS\_$\Sigma$a\_0.02} and \simname{MS\_Tc\_Z0.02}.
Consider the first two models, where the overall trend in accretion history was similar to canonical models with $\Sigma_{\rm a} = 100$~g~cm$^{-2}$ (\simname{MS\_Z0.1} and \simname{MS\_0.02}). 
With a decrease in metallicity, the duration of the outbursting phase decreased, while the individual luminosity bursts became more powerful.
The correlation of the accretion bursts with an increase in inner disk temperature above the critical value indicates that these are MRI-outbursts.
The disk mass for the lower metallicity model was marginally larger at the end of the simulations, with the innermost regions ($< 10$ au) showing a larger difference.

The comparison of \simname{MS\_$\Sigma$a\_Z0.1} and \simname{MS\_$\Sigma$a\_0.02} with their counterparts with canonical active layer thickness shows similar evolution.
The burst phase ended at about 0.42 Myr for \simname{MS\_$\Sigma$a\_Z0.1} which was almost equal to that of \simname{MS\_Z0.1}.
For \simname{MS\_$\Sigma$a\_0.02}, the burst phase lasted 0.22 Myr, which is about 0.05 Myr shorter as compared with \simname{MS\_Z0.02}.
Thus the doubling in the active layer thickness from $\Sigma_{\rm a} = 100$~g~cm$^{-2}$, had only marginal impact on the burst phase. 
With an increased active layer thickness, the overall bottleneck to angular momentum transport in the inner disk was reduced and the disk could accrete gas more efficiently.
This can be inferred by comparing the final stellar mass in \simname{MS\_$\Sigma$a\_Z0.1} and \simname{MS\_$\Sigma$a\_0.02} was larger as compared to their counterparts with canonical active layer thickness.
Most notably, the mass contained within the innermost 10 au was substantially smaller in the case of  models with an increased $\Sigma_{\rm a}$, indicating a large impact on the structure of the inner disk.
Thus we conclude that although the increased $\Sigma_{\rm a}$ at low metallicity has large effects on the structure of the innermost regions of a PPD, the episodic accretion in only marginally affected.

Consider the evolution of \simname{MS\_Tc\_Z0.02} simulation in Figure \ref{fig:1dN2}. 
The accelerated evolution due to the increased cloud core temperature can be inferred from the cloud infall rate in the first panel.
The embedded phase for this simulation lasted only 0.113 Myr as compared to 0.242 Myr for rest of the solar mass models.
This model showed the most extreme shortening of the burst phase, which lasted only 0.13 Myr.
Note that the disk formation also occurred at an earlier time. 
In addition to the shortening, the mass infall rate as well as the base mass accretion rate were larger by a factor of few as compared to rest of the low cloud core temperature models during the early times.
The enhanced accretion rate was responsible for a larger net accretion of material onto the central protostar at the end of the simulation.
As a result, the final stellar mass was larger in this case as compared to \simname{MS\_0.02} simulation, despite  the similar bottleneck of the dead zone.
Near the end of the simulation, the total disk mass was smaller in \simname{MS\_Tc\_Z0.02} as compared to \simname{MS\_0.02}, however, the mass contained within inner 10 au was similar. 
Thus, a substantial fraction of the disk mass was contained within the innermost regions.

\subsection{Lower stellar Mass}
\label{subsec:ML}

In this section we present the results for the lower mass (\simname{ML}) models with an initial core gas mass of 0.576 $M_\odot$, which is half as massive as that of the solar mass models presented earlier.
Such a system would ultimately produce a dwarf star at the boundary of K and M spectral type, with a fully convective interior.
Note that the ratio of kinetic to gravitational energy is kept constant across all of the models.
The mass of the cloud core limits the total mass available for disk formation and a proportionally lower mass disk is seen at all stages of disk evolution.
Lower mass disks are usually less GI unstable, showing suppression in the associated accretion rate variability. 
The accretion via the disk also lasts for a shorter duration in such systems, primarily due to the limited amount of mass reservoir \citep{Vorobyov10}.

\begin{figure*}[t]
\hspace{3.3cm} \simname{ML\_fid}  \hspace{4cm} \simname{ML\_Z0.1}  \hspace{3.5cm} \simname{ML\_Tc\_Z0.02}  \\
\begin{tabular}{l}
\vspace{-0.27cm}\includegraphics[width=0.99\textwidth]{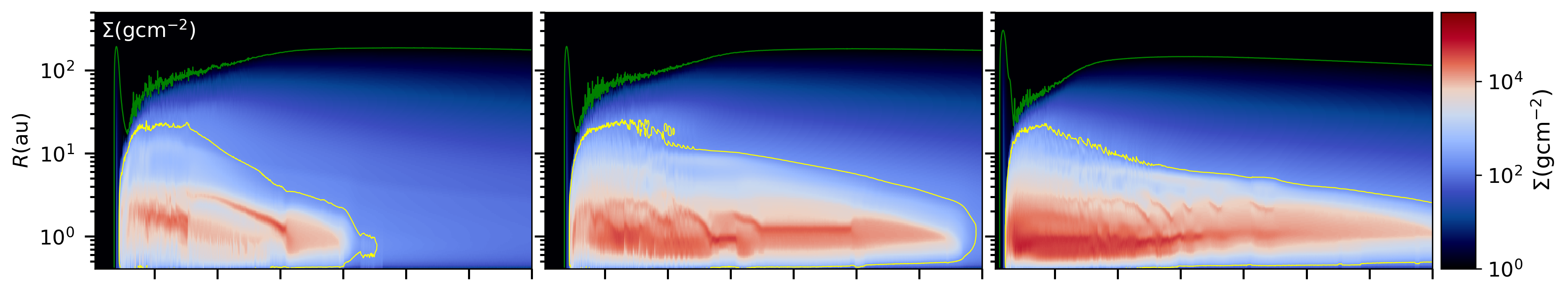} \\
\vspace{-0.27cm}\includegraphics[width=\textwidth]{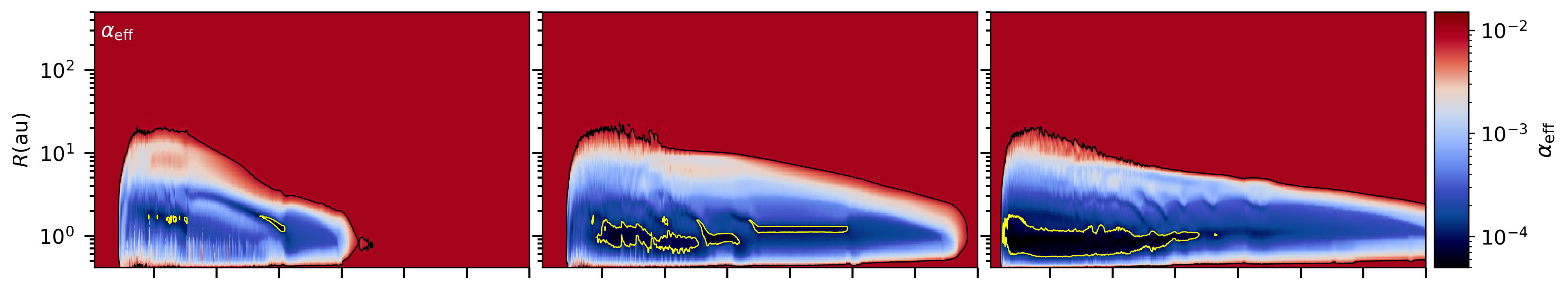} \\
\vspace{-0.27cm}\includegraphics[width=0.995\textwidth]{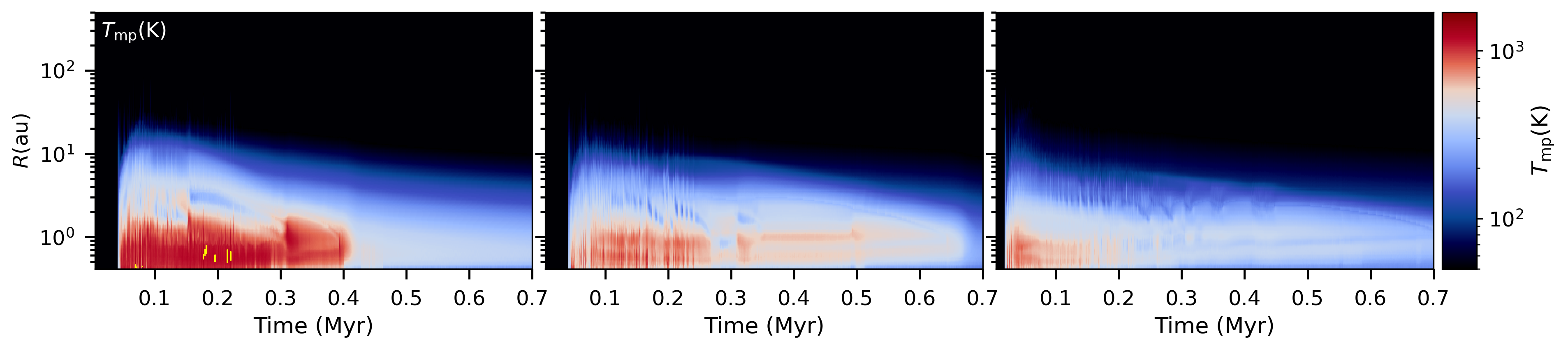} \\  
\end{tabular}
\caption{
The spacetime plots for the three models -- \simname{ML\_fid}, \simname{ML\_Z0.1} and \simname{ML\_Tc\_Z0.02} -- are compared. 
The rows depict the evolution of azimuthally averaged quantities -- $\Sigma$, $\alpha_{\rm eff}$ and $T_{\rm mp}$.
The green and yellow curves in $\Sigma$ show 1 and 200 ${\rm g~cm^{-2}}$ contours, respectively.
The black contour in $\alpha_{\rm eff}$ shows the extent of the dead zone,
while the yellow lines shows $10^{-4}$ contour. 
The yellow lines in the temperature plots show $T_{\rm crit}=1300$ K contours. 
}
\label{fig:spacetimeL1}
\end{figure*}

In Figure \ref{fig:spacetimeL1} we compare the inner disk structure for the three low mass models -- \simname{ML\_fid}, \simname{ML\_Z0.1} and \simname{ML\_Tc\_Z0.02}.
Consider the fiducial low mass model \simname{ML\_fid}.
As compared to the corresponding solar mass model, \simname{MS\_fid}, several key differences can be noticed. 
The most significant difference in terms of the disk structure was that the gaseous rings as well as the dead zone formed were less robust and were sustained for a much shorter time.
After the disk formation at 0.04 Myr, the gas surface density could be maintained above $\Sigma_{\rm a}$ because of the mass loading from the collapsing core. 
This explains the formation of the dead zone and gaseous rings at early times.
With the limited mass contained in the cloud core, the accreted mass was not replenished from the outer disk and the layered structure could not be maintained for an extended period of time.
The temperature of the innermost part remained relatively high as long as the mass transfer rate was maintained and then the disk cooled with time.

Consider the effects of reduced metallicity on the inner disk structure in Figure \ref{fig:spacetimeL1}.
Note that in simulation \simname{ML\_Z0.1} effects of opacity only are considered.
The trends across the simulations are similar to described earlier with the solar mass models.
The outer parts of the disks showed similar behavior while most differences are confined to the innermost regions. 
Comparing spacetime plots of the gas surface density and $\alpha_{\rm eff}$, the gas accumulation increased with decreasing metallicity.
In addition, the extent of the dead zone increased both in radial direction and in time.
As described in Section \ref{subsec:opacity}, the lower gas temperature was caused by more efficient disk cooling in low metallicity environment.
This in turn decreased the kinematic viscosity, contributing to the accumulation of material in the inner disk region.
The inner edge of the dead zone also moved inward due to the lower disk temperatures.

In the last model, \simname{ML\_Tc\_Z0.02}, both the effects of opacity and an increased cloud core temperature were taken into account.
In this case we can see an accelerated evolution, with the disk forming at 0.02 Myr, as opposed to 0.04 Myr for the rest of the lower mass models.
This is expected considering the increased infall velocities resulting from the higher cloud core temperature (see Section \ref{subsec:sigmaTc}).
The dead zone formed in \simname{ML\_Tc\_Z0.02} was more robust as compared to rest of the models as inferred from the $\alpha_{\rm eff}=10^{-4}$ contour lines.
Again, the progressively lower gas temperature as seen in the third row is an indicator of the efficient disk cooling due to reduced opacity.

\begin{figure*}[t]
\hspace{3cm} \simname{ML\_$\Sigma$a\_Z0.1}  \hspace{3.8cm} \simname{ML\_Z0.02}  \hspace{3.5cm} \simname{ML\_Tc\_Z0.02}  \\
\begin{tabular}{r}
\vspace{-0.25cm}\includegraphics[width=\textwidth]{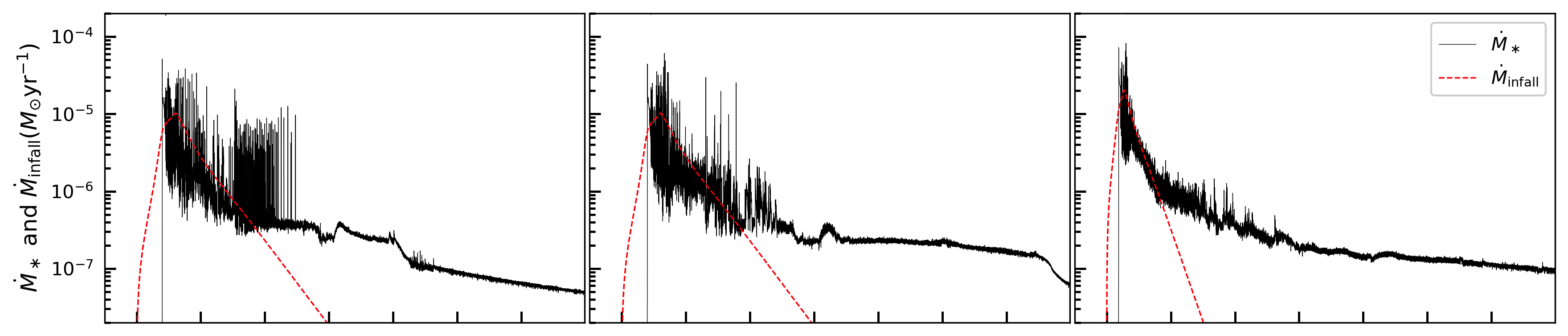} \\
\vspace{-0.25cm}\includegraphics[width=0.988\textwidth]{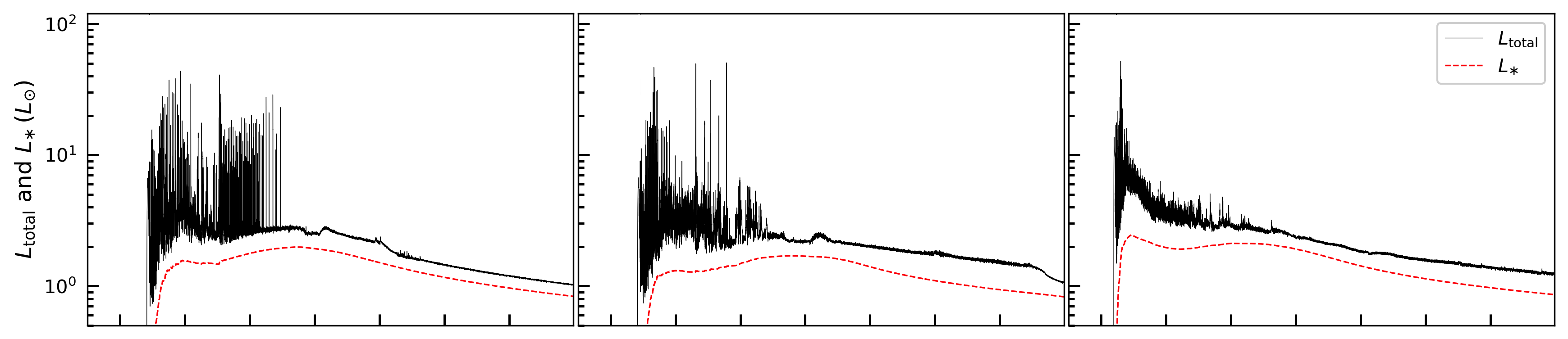} \\
\vspace{-0.25cm}\includegraphics[width=0.996\textwidth]{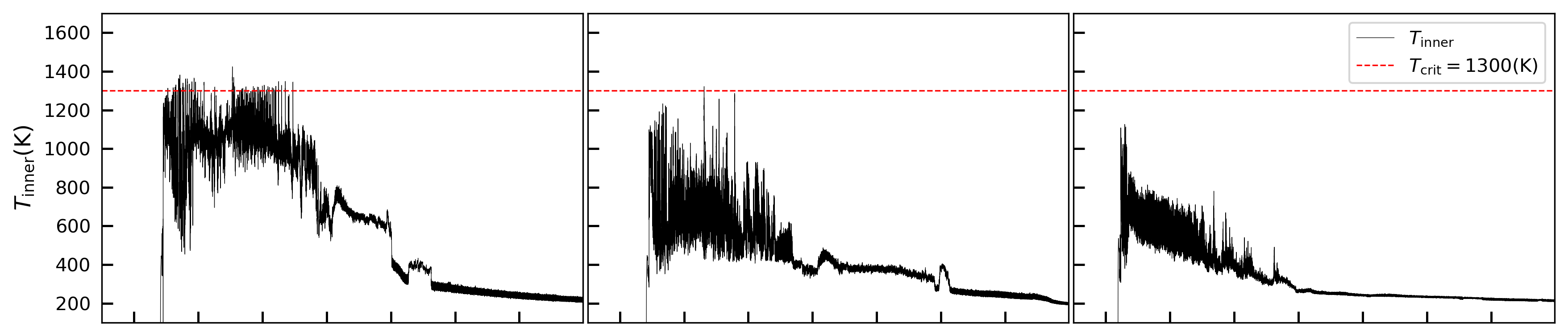} \\  
\includegraphics[width=0.984\textwidth]{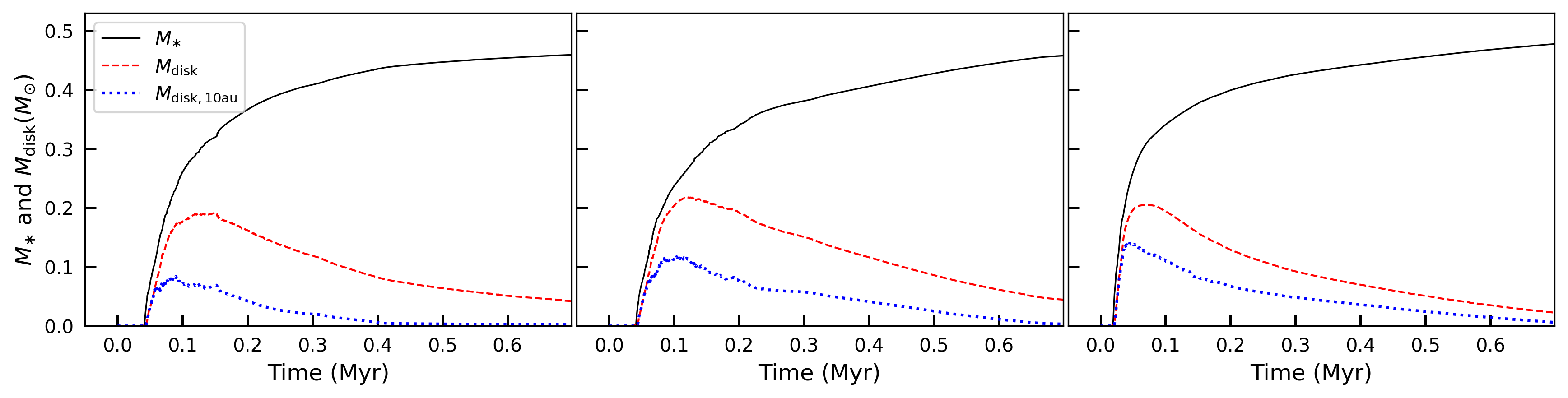} \\
\end{tabular}
\caption{The evolution of time-dependent quantities is compared for the three low mass models -- \simname{ML\_fid}, \simname{ML\_Z0.1} and \simname{ML\_Tc\_Z0.02} 
-- during the entire burst-phase.
The panels show total and stellar luminosities, mean mass accretion rate and cloud core infall rate, temperature at the inner computational boundary, as well as stellar and disk mass, respectively.
}
\label{fig:1dL1}
\end{figure*}

We now focus on the accretion activity in the lower core mass models.
In Figure \ref{fig:1dL1} we compare the time-dependent system properties for the three lower mass simulations.
Consider the fiducial model \simname{ML\_fid}. 
Here, the general features of the mass accretion rate are similar to the higher mass counterpart but limited to a shorter timescale.
The embedded phase in this case lasted about 0.095 Myr, as derived from the mass infall rate. 
The initial phases are dominated by high accretion variability, partly due to the action of GI spirals and partly because of the excursion of the inner edge of the dead zone across the computational domain.
In Section \ref{subsec:opacity} we explained that the inner boundary of the computational domain needs to be small enough to capture the burst phenomenon occurring at the scale of the innermost 1 to 2 au of the disk. 
However, the disk also becomes fully MRI-active in the innermost regions due to thermal ionization, where the disk temperature increases above $T_{\rm crit}$.
This can be seen in the third row, where the midplane temperature at the inner boundary frequently crossed the critical threshold of 1300 K.
For model \simname{ML\_fid}, it was challenging to disentangle the spurious variability due to the interaction of inner edge of the dead zone with the inner boundary of the disk.
However, a few ``bona fide" MRI outbursts, as elaborated later in Section \ref{subsec:burst}, did occur during the early time.
Thus, the MRI bursts were relatively rare in the case of lower mass disks, while the burst phase also lasted for a shorter time, up to about 0.25 Myr.
The evolution of the stellar and disk masses can be observed in the last panel.
Near the end of the simulation, the mass of the protostar approached 0.45 $M_\odot$ while the disk mass continued to decline below 10\% of the stellar mass.

Consider the accretion activity of the low metallicity lower mass simulations -- \simname{ML\_Z0.02} and \simname{ML\_Tc\_Z0.02} -- presented in Figure \ref{fig:1dL1}.
The trends in accretion are similar to those discussed earlier for the solar mass counterparts.
In the lower mass case, the period of accretion variability progressively got shorter with lower metallicity. 
The accelerated evolution for \simname{ML\_Tc\_Z0.02} can be inferred from its higher magnitude and shorter duration mass infall rate, as compared to the lower temperature counterparts.
The inner boundary temperature in \simname{ML\_Z0.1} exceeded $T_{\rm crit}$ several times and triggered MRI outbursts, which can be noticed in concurrent changes in the mass accretion rate and luminosity.
The duration of the burst phase in this case was about 0.18 Myr.
However, the inner boundary temperature in model \simname{ML\_Tc\_Z0.02} did not increase above the critical threshold needed for triggering MRI outbursts, and as a consequence, no luminosity bursts were observed. 
Thus, we conclude that the episodic accretion with the mechanism of MRI bursts was rare in low mass stars as compared to their solar mass counterparts.
In the case of metal depleted lower mass stars, such high luminosity outbursts may be even more uncommon, or completely absent, if the increased cloud core temperature is taken into account.

\subsection{Individual MRI Outburst}
\label{subsec:burst}

\begin{figure}[t]
\centering
\includegraphics[width=0.49\textwidth]{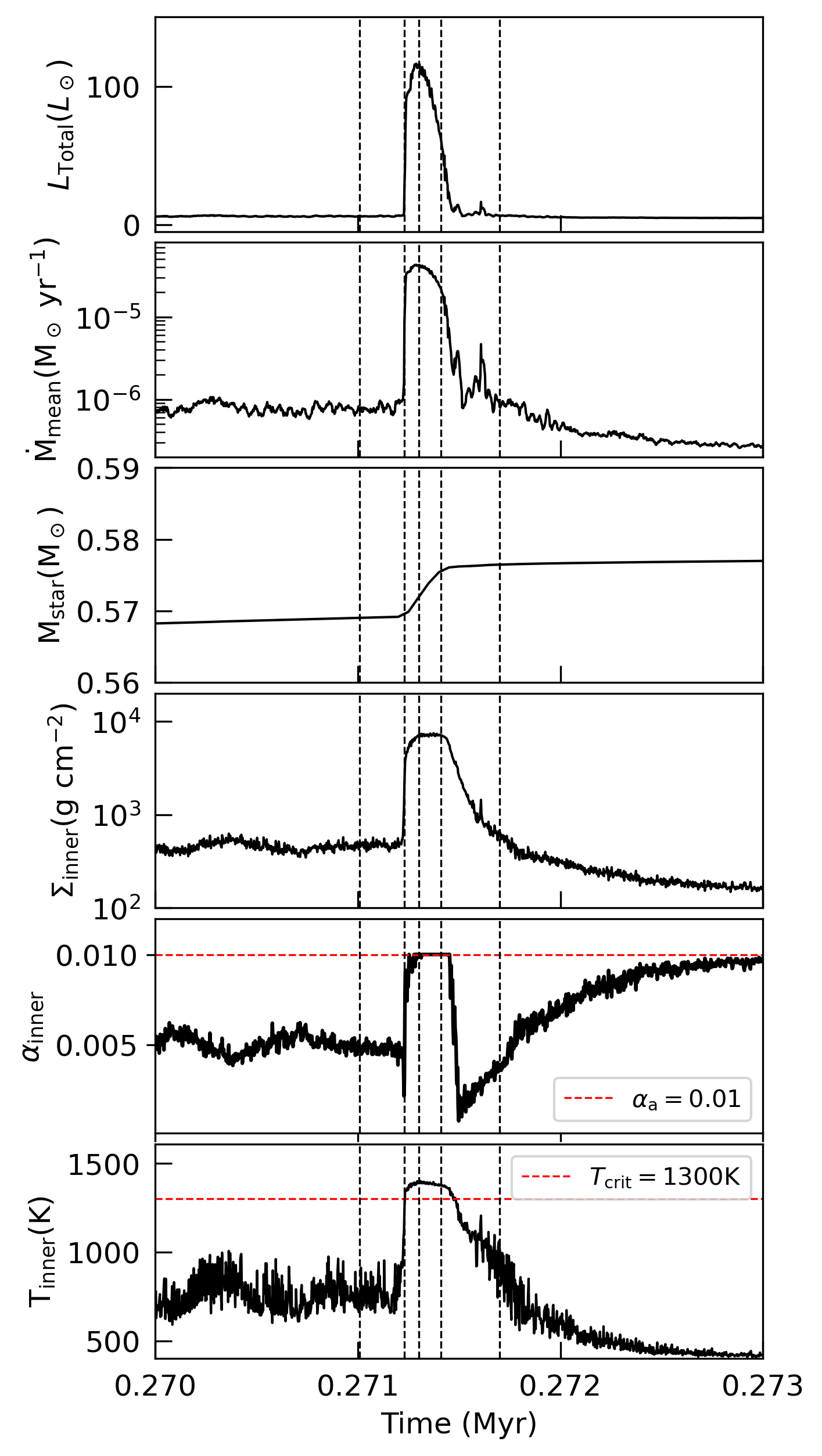}
\caption{
Time dependent system properties (total luminosity, $\dot{M}_{\rm mean}$, stellar mass) as well as surface density, $\alpha_{\rm eff}$ and midplane temperature at the inner boundary during a typical MRI outburst occurring in low metallicity simulation, \simname{MS\_Z0.1}. The vertical dashed lines mark the time corresponding to the 2D plots presented in Figures \ref{fig:mri2D}. The red dashed lines in the last two panels show $\alpha_{\rm bg} = 10^{-2}$ and $T_{\rm crit}=1300$ K, respectively.}
\label{fig:mriLight}
\end{figure}

\begin{figure*}[t]
{\bf \hspace{0.75cm} Pre-outburst \hspace{1cm} $\rightarrow$ \hspace{2.25cm} During outburst  \hspace{2.25cm} $\rightarrow$ \hspace{1cm} Post-outburst}\\
${\color{white}\overbrace{Quit playing games with my heart}}$ $\overbrace{{\color{white}my heart, my heart, I should have known from the start}}$ \\
\begin{tabular}{l}
\vspace{-0.22cm}\includegraphics[width=0.968\textwidth]{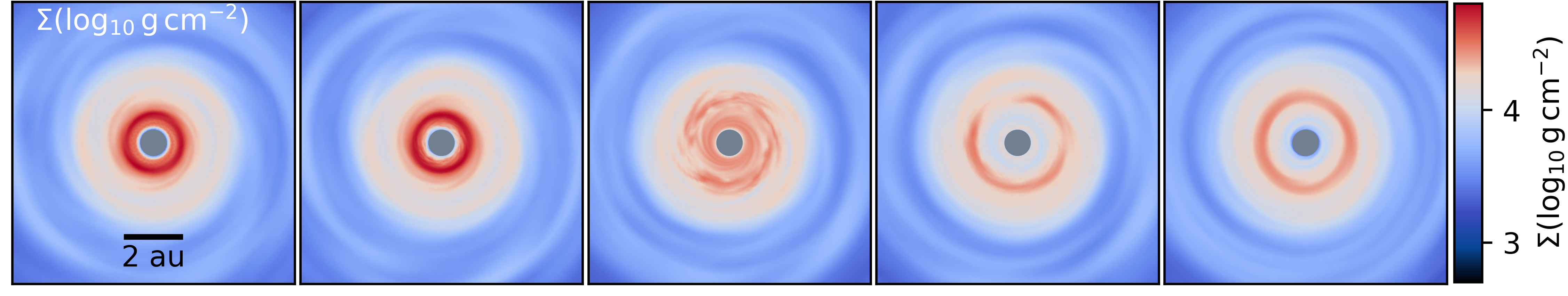} \\ 
\vspace{-0.22cm}\includegraphics[width=\textwidth]{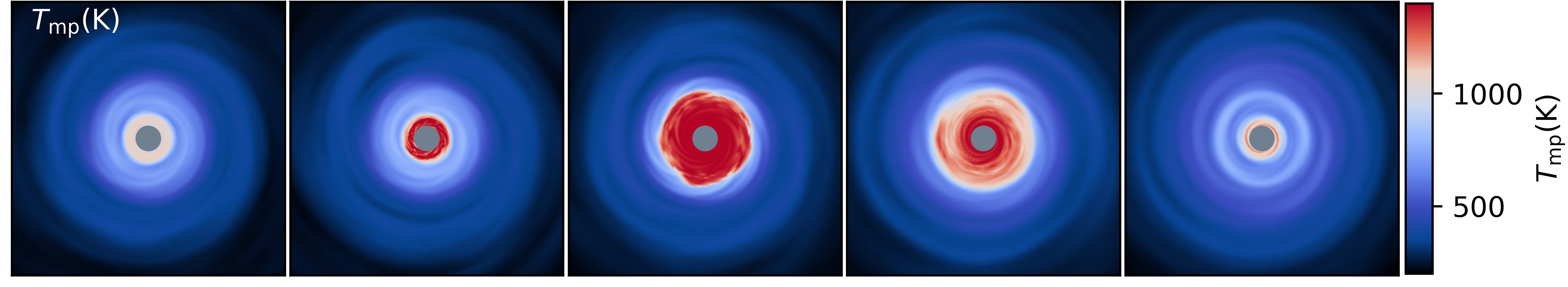} \\
  \includegraphics[width=0.993\textwidth]{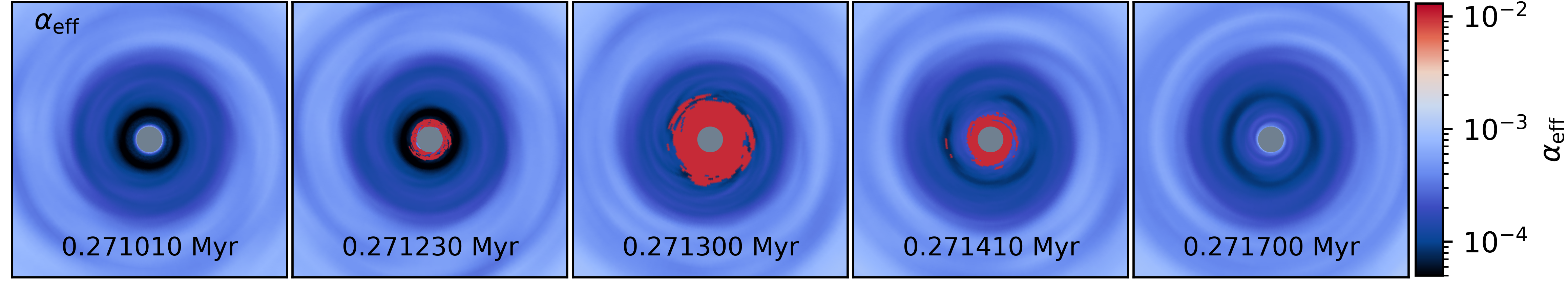} \\
\end{tabular}
\caption{The temporal behavior of the gas surface density, midplane temperature and $\alpha_{\rm eff}$ in the inner $10\times10$ au box of the layered disk, showing the progression of the MRI outburst in the case of low metallicity simulation \simname{MS\_Z0.1}. 
}
\label{fig:mri2D}
\end{figure*}

In this section we present the results of an individual representative example of an MRI outburst in low metallicity environment. 
For direct comparison, the analysis is similar to that performed in \cite{Kadam20} for a solar metallicity model.
We restarted the simulation \simname{MS\_Z0.1} from a suitable checkpoint at about 0.27 Myr and obtained data outputs of 2D fields with a higher time resolution of 20 yr.
The particular representative luminosity burst is highlighted in Figure \ref{fig:1dN1} with an arrow.
Figure \ref{fig:mriLight} shows some of the relevant time-dependent quantities over a period of 3000 yr in the vicinity of this outburst.
Such an MRI outburst empties the inner disk region and causes discontinuity in the spacetime diagram of the surface density, as can be observed in the first row of Figure \ref{fig:spacetimeN1}.
The vertical dashed lines in Figure \ref{fig:mriLight} correspond to the 2D snapshots analyzed later in Figure \ref{fig:mri2D}.
The first panel in Figure \ref{fig:mriLight} shows the  highly asymmetrical light curve profile of the outburst.
The maximum luminosity reached about 118 $L_\odot$ and the total duration was about 200 yr.
The initial rise time of the outburst was relatively fast, about 9 yr, while the decline after the maximum luminosity was much slower as it lasted about 175 yr. 
The duration of the outburst is somewhat smaller than the associated viscous timescale of about 1000 yr, which indicates that the gravitational torques play an important role due to transient non-axisymmetry of the disk.
The mass accretion rate, as shown in the second panel, increased from about $10^{-6}$ to about $4 \times 10^{-5} M_\odot$ yr$^{-1}$.
The accretion rate also showed an asymmetrical, sharp increase and a slower decline.
The stellar mass as shown in the next panel increased by about 0.008 $M_\odot$ over the duration of the outburst, which was about 1.5\% of the mass of the protostar at the time.  
The last three panels show the conditions at the inner boundary across the outburst in terms of azimuthally averaged gas surface density, $\alpha_{\rm eff}$ and midplane temperature.
The sudden increase in $\Sigma_{\rm inner}$ by over an order of magnitude indicates the flow of material across the boundary which was accreted onto the central protostar producing the powerful eruption.
The fundamental reason behind the eruption was the increase of the midplane temperature above the critical value of 1300 K, when the MRI in the disk was triggered (hence the name MRI outburst).
The MRI activation can be seen in the the plot of $\alpha_{\rm inner}$, where it achieves the maximum value of 0.01.
The post-burst values of $\Sigma_{\rm inner}$ as well as $T_{\rm inner}$ were considerably lower as compared to those before the eruption, meaning that the inner disk depleted its mass and became colder.
Since the inner disk was emptied of large amount of material, the $\alpha_{\rm inner}$ increased after the outburst until the gas could be replenished from the outer regions (see Equation \ref{eq:alpha}).

Figure \ref{fig:mri2D} depicts the progression of the outburst in terms of 2D distributions of gas surface density, midplane temperature and effective $\alpha$.
These snapshots (or frames) are marked with vertical dashed lines in Figure \ref{fig:mriLight} and show the innermost $10 \times 10$ au$^2$ region of the disk. 
These particular frames were chosen such that the disk structure could be captured during the rise, at the maximum luminosity and during the decline, as well as before and after the outburst.
The first column shows a quiescent disk before an MRI outburst where the gaseous ring was accumulated at about 1 au from the central protostar.
The second column captures the disk during the sharp increase in mass transfer rate at the beginning of the outburst.
The midplane temperature in the innermost part increased above the critical value, while $\alpha_{\rm eff}$ reached a maximum value of fully active disk, indicating MRI activation.
The third column shows the disk structure near the maximum of the luminosity burst and also corresponds to the maximum extent of the MRI activated region, up to about 1.5 au.
The MRI activated region engulfed the former gas ring, which can be seen almost fully disrupted in this frame.
The accumulated material accreted through the inner boundary due to the increased viscosity, resulting in the large mass transfer rate.
It is clear that this was an inside-out MRI outburst, originating near the inner boundary due to viscous heating and propagating in the outward direction in a snowplough fashion.
The third column shows the disk structure during the decline. 
Only a small region where the temperature remained above the critical value remained fully MRI active at this point.
Note that the disk structure showed significant transient asymmetries during the outburst.
The last column shows a quiescent disk shortly after the outburst. 
As the viscosity channels bulk of the material inward, the angular momentum is transported outward via a relatively small amount of mass. 
This can explain the final low surface density ring surrounding the maximum extent of the MRI active region.

The general properties as well as the mechanism of an individual MRI outburst in a low metallicity environment were same as those in a solar metallicity disk \citep{Kadam20}.
The MRI was triggered at the inner boundary due to increase in temperature above the critical value due to viscous heating in the dead zone, resulting in an inside-out burst.
However, there were a few key differences.
When compared to their solar metallicity counterparts, the outbursts in low metallicity disks were more powerful (in this particular example, 118 versus 75 $L_\odot$) and showed a sharply rising profile (10 versus 100 yr) with a shorter total duration (200 versus 300 yr).
As elaborated in Section \ref{subsec:opacity}, the structure of the innermost region was altered with lowering of the metallicity.
A larger amount of material was accumulated in the gaseous rings, while this material also moved closer to the protostar (see Figure \ref{fig:spacetimeN1}).
As an outburst progressed, the MRI-activated front encountered this mass reservoir within a shorter time as compared to a solar metallicity disk.
This explains the steeper rise-time of the outburst as well as the larger peak mass transfer rate, making the outburst more luminous.
The total duration of the outburst is proportional to the viscous timescale, which was reduced at this shorter radius. 
Note that the total mass accreted in the low metallicity outburst was also larger, approximately 0.008 $M_\odot$, as compared to 0.005 $M_\odot$ in the solar metallicity case \citep{Kadam20}.
In Figure \ref{fig:1dN1} we see a continued trend of increased luminosity of individual bursts with further decrease in metallicity, which can be explained along the same line of reasoning.

\section{Conclusions}
\label{sec:conclusions}

In this study, we conducted numerical experiments using global hydrodynamic simulations of PPD formation in the thin disk limit, with an aim of understanding the effects of metal poor environments -- 10\% and 2\% of the solar metallicity -- on the young disks. 
The primary focus was on the episodic accretion occurring during these early stages of evolution as well as the inner disk structure.
The simulations started with the collapse phase of the initial cloud core and the inner boundary was placed at 0.42 au, so that the MRI bursts occurring at the sub-au scale could be captured.
The dead zone was modelled by an adaptive and effective $\alpha$, which depended on the disk gas surface density as well as midplane temperature.
The low metallicity was modelled by scaling down the gas and dust opacities.
The effects of the increased thickness of the magnetically active surface layer and the temperature of the prestellar cloud core were also studied.
At low metallicities, the general structure and behavior of the accreting disks was similar to that at the solar value. 
The dead zone was formed in the inner approximately 10 au region, with gaseous ring-like structures appearing near the inner edge of the dead zone.
The early evolution was dominated by formation of spirals and GI activity, resulting in variability in the mass accretion rate.
The disks around solar mass stars showed episodic accretion via MRI bursts for a limited time during the evolution of the disk.
We emphasize that the span of the considered spatial scales ($0.4-10 000$ au) and of the evolution period (0.7 Myr) stretch the boundaries of the thin-disk models to their limit.
A simulation in this study typically takes more than a month of computing time on 32 CPU cores and such calculations are prohibitively expensive in full 3D models.

The salient differences between the solar and the lower metallicity PPD formation simulations are summarized as follows.
\begin{itemize}
\item[--] With lower metallicity, the ring-like structures that formed in the dead zone were more robust in terms of accumulated gas as well as low effective $\alpha$ parameter, while the accumulation occurred closer to the central protostar. 
These effects were due to the effective cooling of the disk at low metallicity, which also resulted in low midplane temperatures in general.
\item[--] The mass accumulated in the inner regions of metal poor disks was significantly larger as compared to their solar metallicity counterparts.
This can explain some of the observed trends, e.g., the higher mass accretion rates for low metallicity YSOs at later stages \citep{Spezzi12} and insensitivity of the frequency of short period super-Earths to the metallicity \citep{Petigura18}.
\item[--] The burst phase, when the disk is subject to powerful MRI outbursts, became shorter with decreasing metallicity. This was especially true for the lowest metallicity models if the increased molecular cloud core temperature due to inefficient line cooling was considered. 
Lower inner disk temperatures in metal poor disks also resulted in shortening of the burst phase.
\item[--] The MRI outbursts were rare for a lower mass star of a main sequence mass of $\lessapprox 0.5 M_\odot$, and they were even more unlikely for their low metallicity counterparts, when the increased cloud core temperature was included.
\item[--] An individual MRI outburst in the low metallicity disk was shorter in duration, more luminous, showed a steep rising luminosity curve and accreted more mass, as compared to its solar metallicity counterpart.
\end{itemize}
Although low metallicity PPDs are currently difficult to observe, our predictions may be verified in the future by high-resolution, infrared spectroscopy and imaging with the next generation of large telescopes, e.g., E-ELT and TMT.

Here we mention some of the limitations of this investigation. 
The critical temperature for MRI ignition as well as the thickness of the active layer were considered constant in this study. 
Magnetohydrodynamics equations which incorporate a detailed ionization balance are needed to be solved in order to obtain a better estimate of the disk behavior.
The evolution of the dust component was not considered in this study, which can alter the disk thermodynamics, especially because its accumulation in the regions of the pressure maxima.
The innermost fully MRI-active region (formed due to collisional ionization) could interfere with the inner boundary of the computational domain, producing spurious variations in the mass accretion rate.
This remains a general issue with simulations which terminate at a certain distance away from the central protostar.
The region between the inner boundary and stellar surface is complex and includes several important phenomena such as inner rim of the dust and magnetoshperic accretion.
Despite these uncertainties, the overall picture and trends presented in this study with respect to decreasing opacity should remain valid. 

\section*{Acknowledgements}
{We thank the anonymous referee for constructive comments, which improved the quality of the manuscript.}
This project has received funding from the European Research Council (ERC) under the European Union's Horizon 2020 research and innovation programme under grant agreement No 716155 (SACCRED).
E. Vorobyov acknowledge support from the Austrian Science Fund (FWF) under research grant P31635-N27. The simulations were performed on the Vienna Scientific Cluster (VSC-3 and VSC-4).

\appendix
\vspace{-1cm}
\section{ Comparison across spatial resolution}
\label{appendixA}

When conducting numerical simulations, one usually need to find a balance between computational cost and a grid resolution sufficient to capture the phenomena of interest. 
The implementation of the hydrodynamic equations in \simname{FEoSaD} has been carefully evaluated with several test cases, confirming the ability of our numerical schemes to reproduce the known analytic solutions in the thin-disk limit \citep{VB2006,VB2010}.
As the actual problem of protoplanetary disk formation and its dynamical evolution is fundamentally non-linear in nature, the adequate grid resolution can only be found by trial-and-error.
In this appendix, we present the results of comparative analysis of two identical simulations conducted at different resolutions. 
The fiducial solar-mass model (\simname{MS\_fid}) is compared with its high resolution counterpart, \simname{model1\_T1300\_S100} in \cite{Kadam19}.
Henceforth, the latter model will be termed \simname{MS\_fid\_HR} for convenience. 
The simulations presented in this study, including \simname{MS\_fid}, were all conducted at a resolution of $256\times256$, while \simname{MS\_fid\_HR} model corresponds to twice this resolution, i.e., $512\times512$.
We show that the results of these simulation are congruent and the lower resolution used for the simulations presented in this study was sufficient for capturing the essential aspects of the evolution of the protoplanetary disks. 

Even with the advantages of the thin-disk approximation in our calculations, increasing the grid resolution in \simname{FEoSaD} is computationally demanding for two reasons. 
Doubling the resolution in both R- and $\phi$-direction increases the number of cells by a factor of four.
In addition, because of the cylindrical geometry of the grid, the \cite{CFL1928} condition near the axis of the grid implies more stringent constraints for the numerical stability of the code.
This forces a smaller timestep for the hydrodynamical calculations at higher resolution, making the final computation up to 16 times more expensive.
As a result, the high resolution simulation such as \simname{MS\_fid\_HR} can run on a computer node with 48 cores (\simname{FEoSaD} is OpenMP-parallelized) for several months of real time.  
This longer runtime of the high resolution models was the primary reason for the chosen grid resolution for this study, wherein the entire burst phase needed to be captured.

\begin{figure*}[t]
\hspace{4.5cm} \simname{MS\_fid}  \hspace{6.5cm} \simname{MS\_fid\_HR}\\
\begin{tabular}{l}
\vspace{-0.27cm}\includegraphics[width=0.995\textwidth]{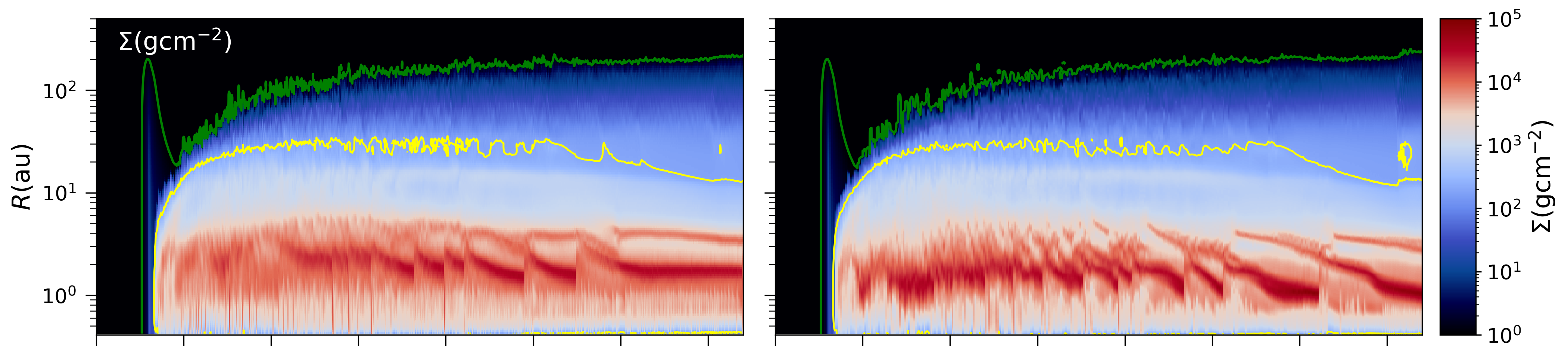} \\
\vspace{-0.27cm}\includegraphics[width=\textwidth]{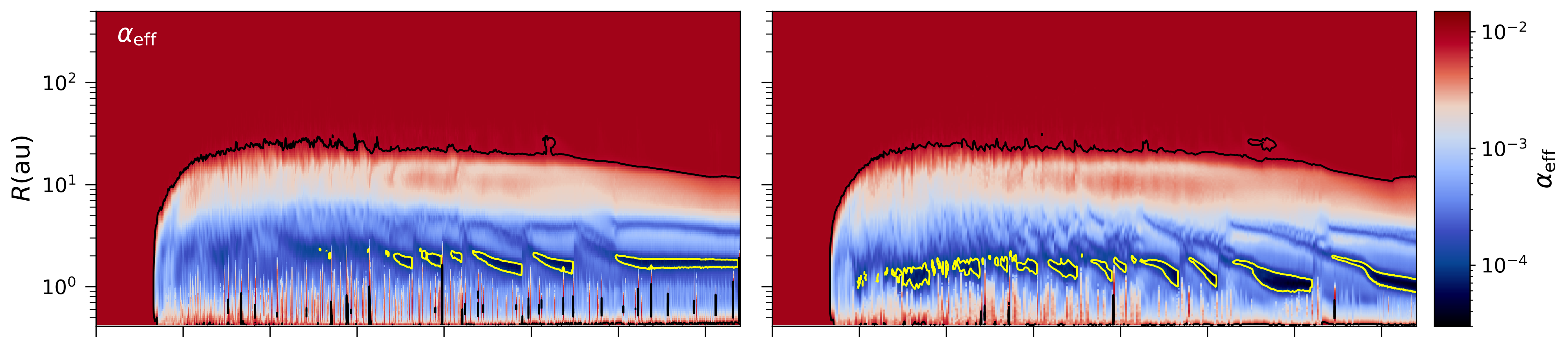} \\
\vspace{-0.27cm}\includegraphics[width=0.995\textwidth]{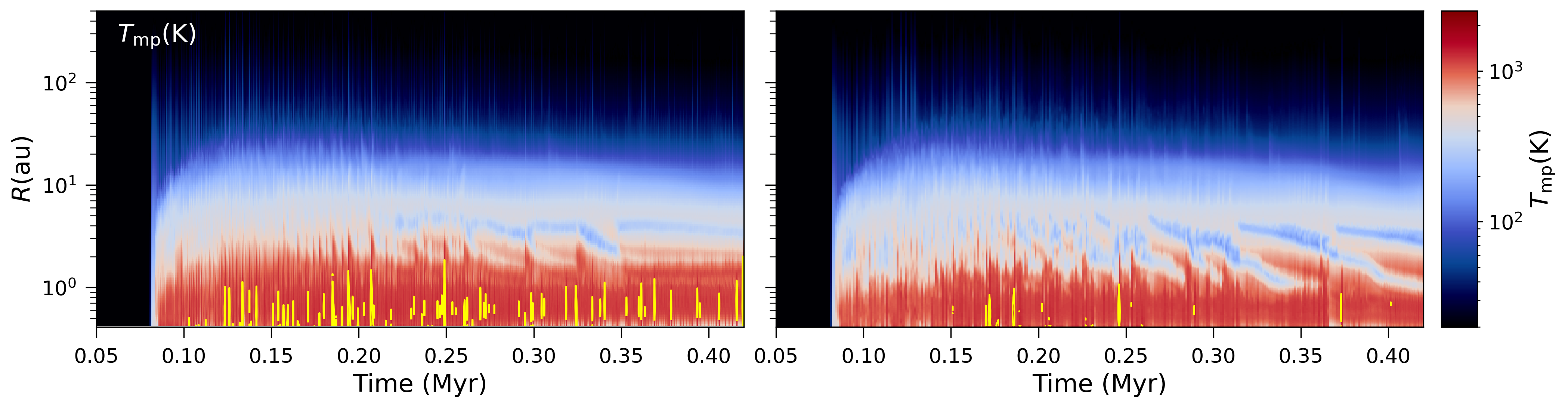} \\  
\end{tabular}
\caption{
The spacetime plots for the fiducial model \simname{ML\_fid} is compared with its higher resolution counterpart \simname{ML\_fid\_HR}. 
The rows depict the evolution of azimuthally averaged quantities -- $\Sigma$, $\alpha_{\rm eff}$ and $T_{\rm mp}$.
The green and yellow curves in $\Sigma$ show 1 and 200 ${\rm g~cm^{-2}}$ contours, respectively.
The black contour in $\alpha_{\rm eff}$ shows the extent of the dead zone,
while the yellow lines shows $10^{-4}$ contour. 
The yellow lines in the temperature plots show $T_{\rm crit}=1300$ K contours. 
}
\label{fig:spacetimeA}
\end{figure*}

\begin{figure*}[t]
\hspace{5cm} \simname{MS\_fid}  \hspace{6.5cm} \simname{MS\_fid\_HR}\\
\begin{tabular}{r}
\vspace{-0.31cm}\includegraphics[width=0.965\textwidth]{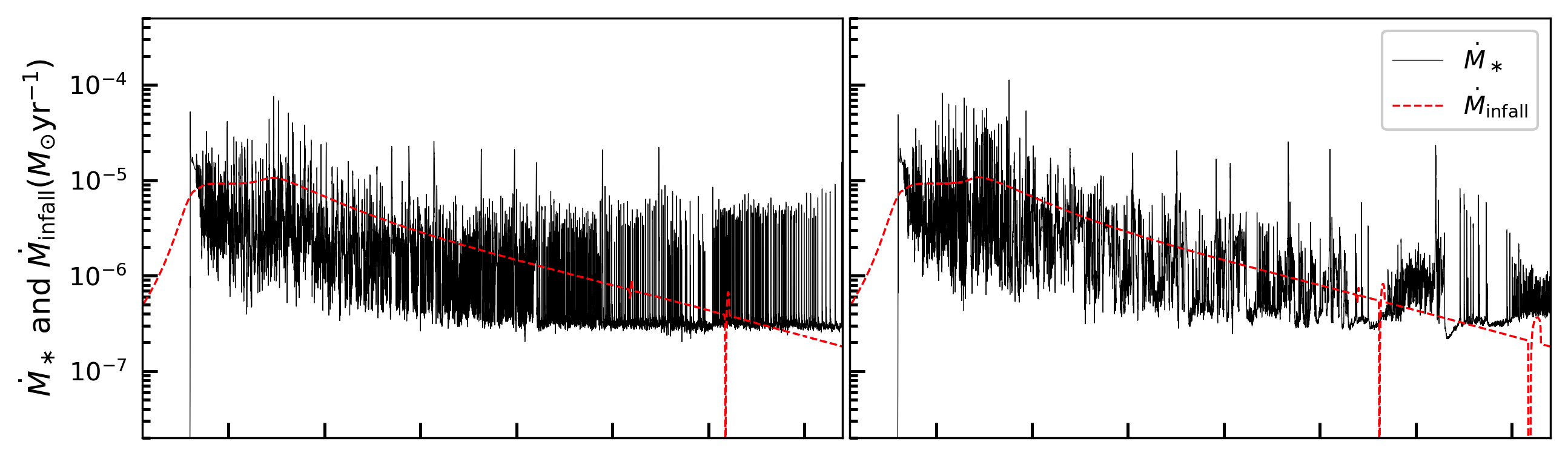} \\
\vspace{-0.31cm}\includegraphics[width=0.95\textwidth]{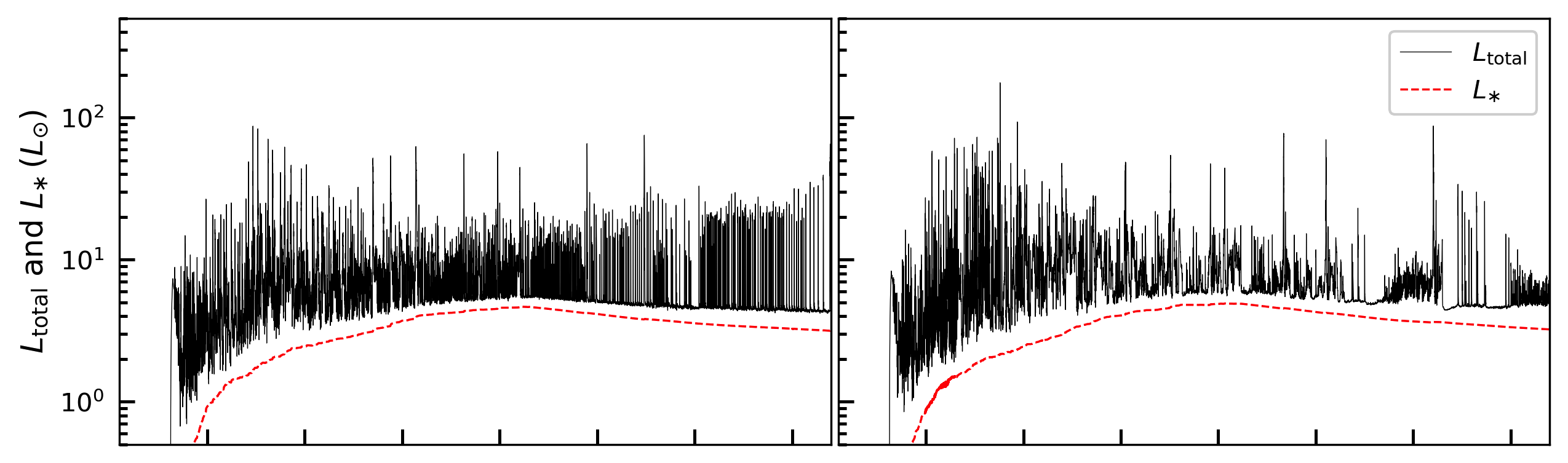} \\
\vspace{-0.31cm}\includegraphics[width=0.96\textwidth]{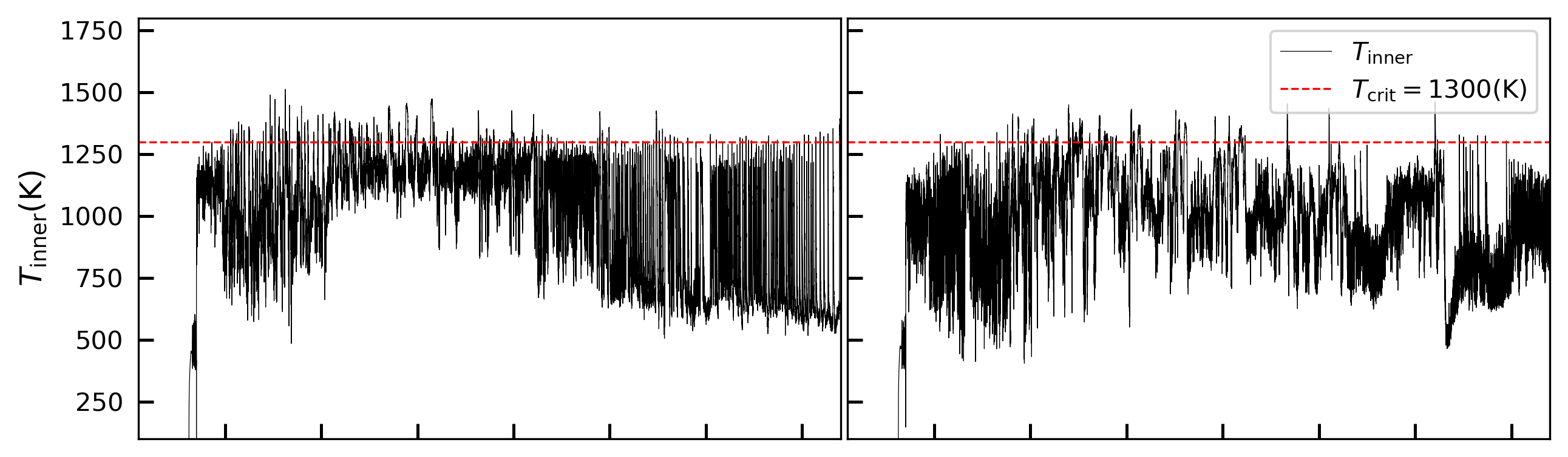} \\  
\includegraphics[width=0.945\textwidth]{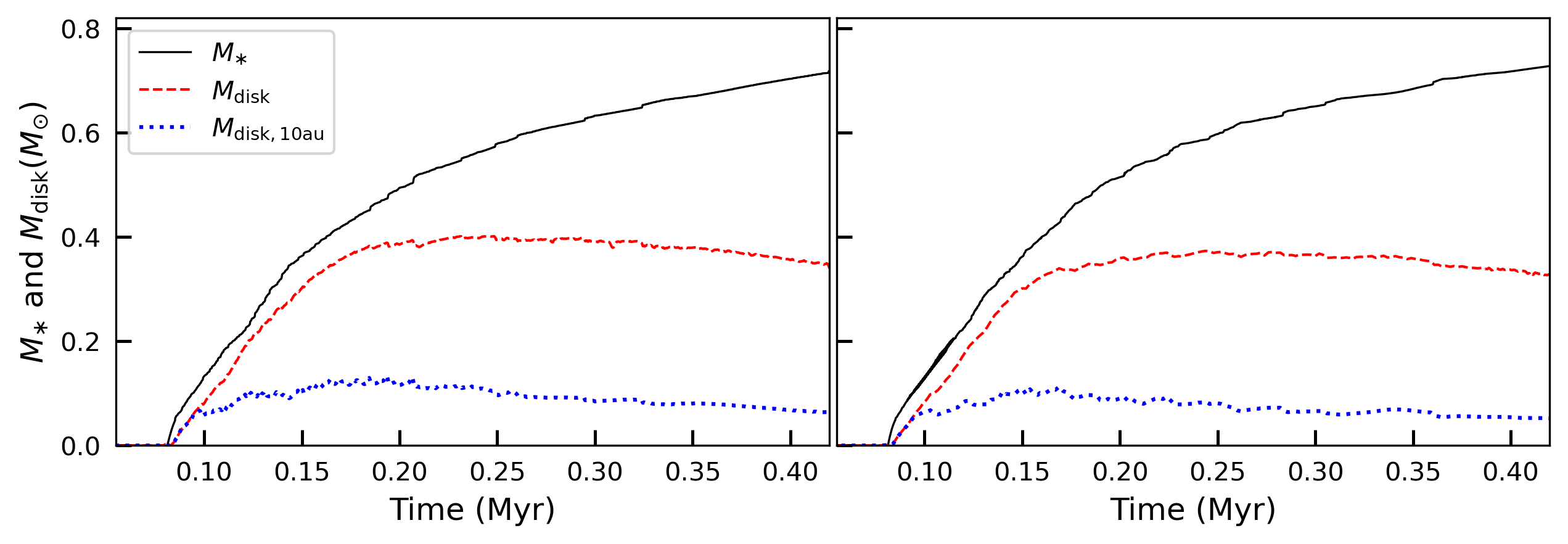} \\
\end{tabular}
\caption{The evolution of time-dependent quantities is compared for the \simname{ML\_fid} model with its higher resolution counterpart \simname{ML\_fid\_HR} -- during part of the burst-phase.
The rows show total and stellar luminosities, mean mass accretion rate and cloud core infall rate, temperature at the inner computational boundary, as well as stellar and disk masses (total and that of the inner disk at $< 10$ au), respectively.
}
\label{fig:1dA}
\end{figure*}

Figure \ref{fig:spacetimeA} compares the inner disk structure of model \simname{MS\_fid} with its high resolution counterpart. 
The overall evolution of the disk structure was similar with respect to all three quantities --  $\Sigma$, $\alpha_{\rm eff}$ and $T_{\rm mp}$.
The disks formed at the same time in both models at about 0.08 Myr.
The outer boundary of the disk ($\Sigma=1$ ${\rm g~cm^{-2}}$ contours), as well as the dead zone (black contours in $\alpha_{\rm eff}$) showed almost identical radial extents at all times.
The rings formed in the inner disk were similar in strength, both in terms of the surface density and viscosity, the latter can be inferred from the $\alpha_{\rm eff}=10^{-4}$ contours.
As elaborated earlier, the discontinuities in the rings typically correspond to MRI bursts, wherein the disk is rapidly accreted through the inner boundary.
The evolution of the disk midplane temperature was also consistent across the two simulations. 
Two key differences between the disk structure were observed in the innermost regions -- the rings in \simname{ML\_fid\_HR} simulation showed much more intricate features, while the temperature crossed the critical threshold less frequently.
The radial extent of the computational domain in both the models was identical, extending from $0.4167$ au to $10210$ au.
The highest grid resolution in the radial direction was $1.6 \times 10^{-2}$ au at the innermost cell of the disk for \simname{MS\_fid}, as compared to $8.2 \times 10^{-3}$ au for \simname{MS\_fid\_HR} model.
Thus, the high resolution model was able to spatially resolve the inner disk much better and
the less frequent temperature excursions in \simname{MS\_fid\_HR} model may also be a result of the improved resolution.
However, the general picture with respect to the disk structure and its evolution remained unchanged for the two models.

In Figure \ref{fig:1dA} we compare the low and high resolution models with respect to time-dependent properties relevant to episodic accretion.
The overall behavior for all four panels was very similar.
Note that the simulations are compared up to 0.42 Myr of evolution and not during the entire burst phase.
As explained in Section \ref{subsec:opacity}, the mass accretion rate was a superposition of the three modes of variability -- excursion of the inner edge of the dead zone across the computational boundary, MRI bursts and vigorous GI activity.
The temperature at the inner boundary crossed the MRI activation threshold, with the larger amplitude changes corresponding to the MRI outbursts. 
The evolution of the stellar and disk masses was almost identical across the resolutions for the two models.
One notable difference between the two simulations can be seen in the behavior of mass accretion rate, and hence the total luminosity.
In between the larger MRI bursts, the background variability was diminished with the increase in resolution.
This was also reflected in the evolution of $T_{\rm inner}$, where it crossed the critical threshold of 1300 K less frequently in between the larger amplitude MRI outbursts for \simname{MS\_fid\_HR} model.
As part of the small-scale variability was caused by the movement of the inner edge of the dead zone across the computational boundary, it may be improved with the better resolved innermost regions.
The overall evolution remained congruent for the two simulations, and we conclude that the resolution chosen for this study was sufficient to capture both the inner disk structure and the outbursting behavior of the systems.

\end{document}